Fractional-Monolayer 2D-GaN/AlN Structures: Growth Kinetics and UVC-emitter Applications

*Valentin Jmerik*, Dmitrii Nechaev, Evgenii Evropeitsev, Evgenii Roginskii, Maria Yagovkina, Prokhor Alekseev, Alexey Semenov, Vladimir Kozlovsky, Mikhail Zverev, Nikita Gamov, Tao Wang, Xinqiang Wang, Tatiana Shubina, Alexey Toropov,*

V.N. Jmerik, D.V. Nechaev, E.A. Evropeitsev, E. M. Roginskii, A. Semenov,
M. A. Yagovkina, P. A. Alekseev, T.V. Shubina, A. A. Toropov
Centre of Nanoheterostructure Physics, Ioffe Institute, 26 Politekhnicheskaya,
Saint Petersburg, 194021, Russia
E-mail: jmerik@pls.ioffe.ru

V. I. Kozlovsky, M. M. Zverev, N. A. Gamov
[2]Lebedev Physical Institute, Leninsky Avenue 53, Moscow 119991, Russia;

Tao Wang, Xinqiang Wang
[3]State Key Laboratory for Mesoscopic Physics and Frontiers Science Center for Nanooptoelectronics, School of Physics, Peking University, Beijing 100871, China

Funding: National Key R&D Program Intergovernmental International Science and Technology Innovation Cooperation, Project №2022YFE0140100

## Abstract

The paper reports on fundamental properties of the GaN/AlN quantum wells (QWs) with nominal subcritical thicknesses of 0.75–2 monolayers (MLs). They are grown by plasma-activated molecular beam epitaxy, varying either the nominal thickness or the gallium-to-nitrogen flux ratio. In situ monitoring reveals difference in 2D nucleation and step-flow growth modes of the QWs. The emission characteristics of QWs with integer thicknesses of 1 and 2 MLs depend weakly on the growth mechanism. In contrast, the intensity and spectral position luminescence of QWs with fractional-ML thicknesses are determined by the growth mechanism. Using ab initio calculations, a phenomenological model is proposed that describes fractional-ML QWs either as arrays of 2D quantum disks or as arrays of 2D quantum ribbons, in cases where 2D nucleation or step-flow growth mechanisms predominate, respectively. This model is generally consistent with experimental data on photo- and cathodoluminescence of heterostructures with multiple (250×) GaN/AlN QWs. These heterostructures, when pumped by electron beam at an energy 12.5 keV with a maximum pulse current of 2 A, exhibit linear current dependences of optical peak powers up to 1 and 37 W for wavelengths of 228 and 256 nm, respectively, making them promising for use as powerful ultraviolet-C emitters.

## 1. Introduction

Throughout the development of semiconductor optoelectronics, there has been a continuous miniaturization of device designs. One of the pioneers of semiconductor lasers, Alferov, noted in his Nobel lecture[1] that progress in this field in the 1970s and 80s was ensured by reducing the thickness of the active (light-emitting) regions from 500 nm in the first laser structures with a double heterojunction to < 10 nm in heterostructures with Quantum Wells (QW)[2,3] and Quantum Dots (QD)[4,5]. The spatial localization of charge carriers in the objects with a smaller band gap compared to the surrounding barrier layers mitigates the decrease in radiative recombination efficiency in these structures with increasing temperature and defect concentration.

A wide range of quantum-dimensional objects are formed using several basic growth mechanisms of the heterostructures classified by Bauer in 1958.[6] According to this classification, the two-dimensional (2D) nucleation growth mechanism, named after its discoverers as Frank-van der Merwe (F-M) growth, realized in crystallographically matched or slightly mismatched (<1%) heterostructures, ensures the epitaxial growth of QWs with a typical thickness of several nanometers. Otherwise, the three-dimensional (3D) Volmer-Weber growth mechanism in the highly mismatched heterostructures (>10%) results in formation of relatively large 3D objects. The appearance of nanoscale QDs becomes possible due to the so-called Stranski-Krastanov growth mechanism in heterostructures with moderate crystallographic mismatch.[7] In this two-stage 2D/3D growth, the initial accumulation of elastic energy in 2D films after reaching a critical thickness with typical values of 1–2 monolayers (ML) leads to a transition to 3D growth of QDs with characteristic sizes in the nanometer range.

The first GaN QDs in an AlN matrix were grown by Daudin's group in 1997 using the Stranski-Krastanov growth mechanism in plasma-activated MBE (PA MBE).[8–11] Various GaN/Al(Ga)N QDs were grown by several groups using the same approach,[12–17] ammonia MBE ($NH_3$-MBE),[18–20] as well as metalorganic chemical vapor deposition (MOCVD).[21] Droplet epitaxy was developed to form GaN QDs using either $NH_3$-MBE[22] or MOCVD.[23] The typical thickness of these QDs of 1–7 nm ensures quantum confinement of charge carriers, since their photoluminescence (PL) or cathodoluminescence (CL) spectra exhibit UV emission with energies exceeding the GaN band gap of 3.4 eV. In most studies, the radiation energy does not exceed 4 eV, and only in some studies[10,20,23] the observed higher values of up to ~5 eV were attributed to the ML-thick wetting GaN layer. The most important advantage of GaN/Al(Ga)N QDs over AlGaN-based QWs is their increased quantum efficiency due to charge carrier localization.[16] However, in the wurtzite QW and QD heterostructures, large differences in spontaneous and piezoelectric polarizations between the boundary layers[24,25] cause the so-called Quantum-Confined Stark Effect (QCSE), which consists of the generation of an internal electric field leading to detrimental spatial separation of recombining charge carriers and a red shift of the output emission.[9,15,16,19]

One of the most important areas of modern physics and technology of low-dimensional structures is the development of 2D-layers of various materials. First, graphene discovered in 2004 by Geim and Novoselov,[26] is actively used both in van der Waals epitaxy[27] and for the development of new types of electronic and optoelectronic devices.[28] Furthermore, 2D-dichalcogenides, disulfides, and III-nitrides, have been intensively explored.[29] In addition to standard 2D-materials with planar topology, there is a certain interest in 2D-nanoribbons of various materials with a width of $< 50$ nm, discovered in 1996 by Fugita,[30] and the physicochemical properties of which differ significantly from those of their planar counterparts.[31] Although most studies are devoted to graphene 2D-nanoribbons, there are studies on other materials, including GaN, AlN, and related heterostructures.[32]

In addition to the research areas described above, Ploog's group demonstrated in the early 1990s the capabilities of MBE to grow InAs islands with nominally subcritical thicknesses on the GaAs layers grown on vicinal substrates.[33] These objects, called sub-ML or fractional ML QDs, have been realized in various semiconductors and have attracted attention from both their fundamental optical properties and the development of new types of photodetectors, light-emitting devices, and solar cells.[34–36]

The trend of thinning active regions in light-emitting devices to the ML level has also extended to III-N-based devices. In the early 1990s, Khan proposed replacing the AlGaN layers in UV optoelectronic devices with ultra-short-period (several MLs-range with a GaN ML thickness of 0.259 nm)) GaN/AlN superlattices (SLs), the so-called digital alloys.[37] After first demonstrating this approach using MOCVD, Nikishin's group successfully fabricated digital alloy-based UVC LEDs using $NH_3$-MBE in the early 2000s.[38,39]

In 2008, a sub-ML discrete PA MBE for fabrication of UV-emitting heterostructures with active regions in the form of ultrathin (<1ML) GaN QWs in AlGaN barrier layers was proposed by our group.[40,41] Then, theoretical studies by Kamiya et al.[42] and Strak et al.,[43] as well as experimental work by Taniyasu et al.[44] demonstrated the capabilities of ML-thick GaN/AlN QWs to significantly suppress QCSE as well as completely eliminate the detrimental TE/TM switching of the UVC output polarization mode in Al-rich AlGaN. Over the past decade, UVC-emitting ML-thick GaN/Al(Ga)N QW structures have been grown using both PA MBE[45–58] and MOCVD.[59–63] The general tendency to decrease the emission wavelength to a minimum of ~220–230 nm with a decrease in the nominal QW's thickness to 1 ML or less was noted in thematic reviews.[64,65] However, a systematic comparison of the radiative properties for different ML-thick QWs configurations remain lacking.

In this paper, we study both the 2D-nucleation and the step-flow growth mechanisms of GaN QWs with subcritical thicknesses (up to 2 ML) on ML-stepped atomically smooth AlN surfaces under various PA MBE growth conditions. A comprehensive study of such multiple QWs (MQW) structures, using X-ray diffraction (XRD) and X-ray reflectance (XRR) at grazing angles, atomic force microscopy (AFM) and scanning transmission electron microscopy (STEM) confirm not only their high structural quality, but also the controlling the nominal QW's thickness with an accuracy of 0.25 ML. In parallel, ab initio calculations of the electronic structures of GaN/AlN QWs with various ML and bilayer (BL) well configurations are performed. Based on the results of these calculations and analysis of the PL and CL spectra of the MQW structures with various parameters, a phenomenological model is developed that describes the relationship between the 2D-growth modes of these MQW structures with the formation of different quantum-sized objects, emitting in the UVC spectral range of 228 – 260 nm with various intensity. As a result, electron-beam-pumped UVC emitters with output peak optical powers of up to several tens of watts are demonstrated.

## 2. Main 2D-growth modes of monolayer-thick GaN/AlN quantum wells

### 2.1. Critical thickness of 2D-3D transition of GaN growth mode on AlN barrier layers

Two series of $250\times\{GaN_m/AlN_n\}$ structures (where $m = 0.75 – 2$, $n = 15$ – the thicknesses of the GaN QWs and AlN layers, respectively, in units of the thicknesses of the atomic layers of these compounds, respectively, equal to half the *c*-lattice constants of the unit cells $1/2c_{GaN(AlN)} = 0.259(0.249)$nm) were grown by PA MBE in this work. In the first, so-called "*Thickness-series*", designated as *T-m*, the *m* values varied from 0.75 to 2 ML in 0.25 ML increments at a constant ratio of the Ga and activated nitrogen fluxes $\phi_{Ga}/\phi_{N2*} = 1.5$. In other "*Flux-series*", designated as $F\text{-}(\phi_{Ga}/\phi_{N2*})$, the $\phi_{Ga}/\phi_{N2*}$ varied from 0.7 to 2.1 at a constant $m = 1.5$ ML. 15-ML-thick AlN layers terminated all structures. In addition, two control 1.5-ML-thick GaN QWs were grown at different $\phi_{Ga}/\phi_{N2*} = 1.1,2.1$ on top of the *T*-series structures with $m = 1.5$. For details of the PA MBE growth, see the experimental section.

Figure 1a, S1, Supporting Information shows the AFM images of all samples from the *T*- and *F*-series, which almost all have the same surface topography consisting of atomically smooth ascending hexagonal terraces with a typical width $\omega_T$ = 30–40 nm and a step height of 1–2 ML (Figure S2a,b, Supporting Information). All structures exhibited continuously streaky

reflection high-energy electron diffraction (RHEED) patterns, and an average root mean square (RMS) surface roughnesses estimated from AFM images ($RMS^{AFM}$) over an area of 1×1 μm² was 0.30±0.07 nm in most structures. Only one *T-2* ML structure with the thickest QWs of 2ML showed a relatively sharp increase in $RMS^{AFM}$ to 0.59 ML with almost complete disappearance of atomic steps (Figure S2c, Supporting Information). AFM images of the *T*- and *F*-series structures obtained over a larger scanning area of 5 × 5 μm$^2$ (Figure S1e–j, Supporting Information) showed a relatively slight increase in their $RMS^{AFM}$ to 0.46±0.07 nm in most structures and to ~0.88 nm in *T-2* ML structure. This indicates the stable 2D-growth mechanism throughout 250×periods of the MQW structures without any signs of step bunching. These observations are consistent with the data of Kaufmann et al.[66] who demonstrated the advantage of PA MBE under metal-rich conditions for suppressing the Ehrlich-Schwobel barrier,[68,69] which serves as one of the driving forces of step bunching.

Figure S3, Supporting Information shows a sharp increase in the intensity of specular reflex of RHEED (hereinafter referred to as RHEED intensity ($I_{RHEED}$)) during the growth of thicker GaN layers (>2 ML). The nominal QW's thickness corresponding to this transition depended on the growth conditions. During the growth at $\phi_{Ga}/\phi_{N2*} = 1.1$, this transition was observed at $m = 2.4$ ML (Figure S3a, Supporting Information), whereas when growing GaN under highly metal-enriched conditions with $\phi_{Ga}/\phi_{N2*} = 2.1$, only an initial decrease in $I_{RHEED}$ was observed, reaching a constant minimum level that was maintained until the end of growth at $m = 20$ ML, after which it increased (Figure S3b, Supporting Information). These results are consistent with the data of other authors who studied the 2D-3D transition during GaN PA MBE growth on AlN,[11,14] and with the general theory of critical thicknesses in various heterostructures developed by Marriet[67] taking into account the possible influence of metallic adlayers on the surface energy of epilayers in heterostructures. The growth GaN layers with supercritical thickness on AlN will be described in more detail in a separate article.

Continuous 2D-growth of MQWs was also achieved by depositing metallic Al-BLs on the GaN QW surfaces prior to the growth of each AlN barrier layer.[52] This, combined with the slightly Al-enriched growth conditions of these layers, provided enhanced surface mobility of adatoms throughout their growth. Atomically smooth AlN terraces were exposed to an activated nitrogen flux prior to the growth of each QW to bind excess Al. The identical topography of almost all MQW structures indicates the dominant role of AlN growth conditions in producing indistinguishable AlN surfaces prior to the growth of each QW and points to the need for a more thorough study of the QW's growth process to explain their different structural and optical properties, which is the subject of the following sections.

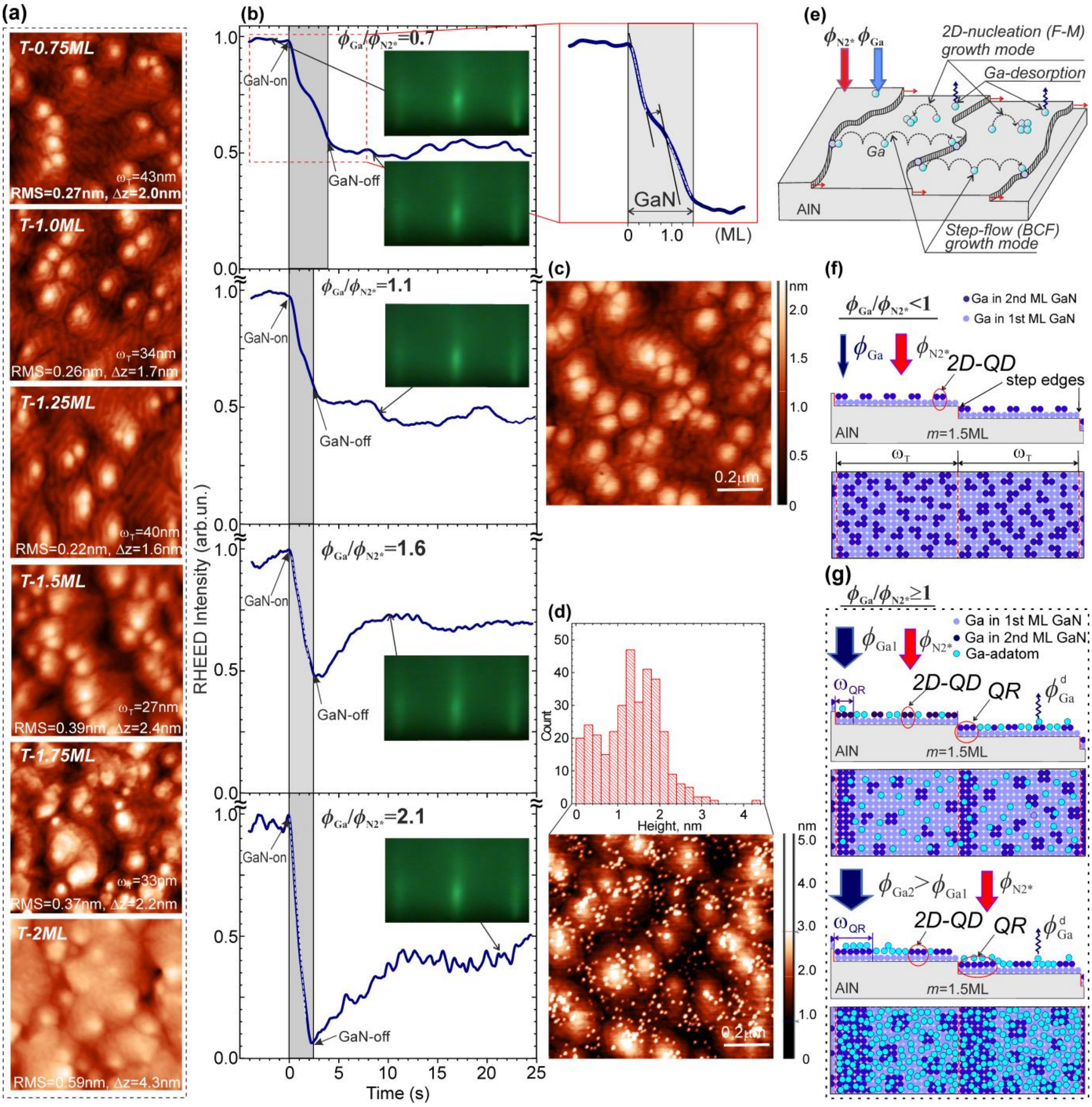

**Figure 1.** (a) AFM images ($1\times1\ \mu m^2$) of the top AlN surfaces of MQW structures from the *T*-series ($\Delta z$ is the maximum height, for $\omega_T$ and RMS see text). (b) Time evolution of $I_{RHEED}$ during growth of 1.5-ML-thick GaN QWs on AlN layers in the *F*-series. The insets show RHEED patterns at the indicated times. AFM images of control GaN QWs with the same $m = 1.5$ML grown on top of $250\times\{GaN_{1.5}/AlN_{15}\}$ MQW structures at different values of $\phi_{Ga}/\phi_{N2*} = 1.1$ (c) and $\phi_{Ga}/\phi_{N2*} = 2.1$ (d). (e) Schematic illustration of the growth of a GaN ML on a stepped AlN layer according to the 2D-nucleation (F-M) and step-flow growth (BCF) mechanisms. Conventional surface diagrams illustrating the formation of 2D-QDs and QRs in the fractional-ML GaN QWs ($1 < m < 2$) grown on stepped AlN surfaces at different $\phi_{Ga}/\phi_{N2*}$ values: <1 (f), ≥1 at various $\phi_{Ga}$ (g). (for $\omega_{QR}$, $\phi_{Ga}^{d}$ see text in section).

### 2.2 Main 2D-growth modes of fractional-monolayer GaN quantum wells

Figure 1b shows the $I_{\mathrm{RHEED}}$ changes during the growth of QWs in *F*-series structures. Although all structures continuously exhibited a streaky RHEED pattern consistent with the maintenance of a 2D-growth mechanism, the $I_{\mathrm{RHEED}}$ behavior was highly dependent on the QW growth conditions. According to Ichimiya and Cohen[70], the evolution of $I_{\mathrm{RHEED}}$ depends on the changes in the chemical composition and surface topography of growing GaN/AlN MQW structures, and several processes that mediate these processes will be analyzed below.

Firstly, the changes in $I_{\mathrm{RHEED}}$ are driven by the change in the average composition of the near-surface heterostructure region, the thickness of which corresponds to the penetration depth of the electron (e-) beam – 1 – 2 MLs at an e-energy of 30 keV and an incidence angle of ~1°. Therefore, the overall decrease in the $I_{\mathrm{RHEED}}$ observed during the growth of all GaN/AlN QWs follows from its dependence $I_{\mathrm{RHEED}} \sim 1/P_{\mathrm{III\text{-}N}}$, where $P_{\mathrm{III\text{-}N}}$ is the average internal potential of the near-surface material, and $P_{\mathrm{AlN(GaN)}} = 14.23(16.82)$ eV, respectively.[71] The same general decreasing shape of the $I_{\mathrm{RHEED}}$ changes for all structures indicates the determining role of this compositional factor.

Furthermore, additional $I_{\mathrm{RHEED}}$ variations can be caused by changes in the surface topography during layer-by-layer growth of GaN in the so-called 2D-nucleation (F-M) growth mode, which forms 2D-GaN ML-thick atomic islands (clusters). The vertical side walls of these islands act as diffuse electron scattering surfaces, and their perimeter varies nonlinearly and periodically with a period corresponding to the complete filling of one ML. The resulting RHEED oscillations have been thoroughly studied for both conventional III-V compounds[70,72] and III-group nitrides.[8–10] Therefore, the small kinks in the $I_{\mathrm{RHEED}}$ linear change observed during the growth under nitrogen-rich ($\phi_{\mathrm{Ga}}/\phi_{\mathrm{N2*}} = 0.7$) and slightly Ga-rich ($\phi_{\mathrm{Ga}}/\phi_{\mathrm{N2*}} \leq 1.1$ and $\phi_{\mathrm{Ga}} < \phi_{\mathrm{N2*}} + \phi_{\mathrm{Ga}}^{\mathrm{d}}$, where $\phi_{\mathrm{Ga}}^{\mathrm{d}}$ is the Ga-desorption flux[73]) conditions can be explained by the formation of atomic GaN 2D-islands during filling of the first GaN ML on atomically smooth AlN terraces in the F-M growth mode. This hypothesis is supported by the observation of an inflection point at a time corresponding to a nominal filling of approximately half a ML (inset in the upper Figure 1b), at which the perimeter of the diffusing sidewall edges of GaN 2D-clusters reaches its maximum value.

Comparison of the $I_{\mathrm{RHEED}}$ behavior during GaN growth at different values of $\phi_{\mathrm{Ga}}/\phi_{\mathrm{N2*}}$ shows a more pronounced kink at lower values of this ratio and its almost complete disappearance during GaN growth under moderately and highly metal-enriched conditions $\phi_{\mathrm{Ga}}/\phi_{\mathrm{N2*}} \geq 1.6$, which corresponds to $\phi_{\mathrm{Ga}} \geq \phi_{\mathrm{N2*}} + \phi_{\mathrm{Ga}}^{\mathrm{d}}$. According to the generally accepted theory of Northrup and Neugebauer,[74] the formation of a Ga-adlayer at Ga-rich conditions of PA MBE increases the surface mobility of adatoms, which ensures a transition from the F-M growth mode to the 2D-step-flow mode in accordance with Burton-Cabrera-Frank (BCF) theory[75]) with limited island nucleation and predominant diffusion of adatoms towards the step-edges. In this growth mode the stepped surface topography does not change,[72] and the influence of this non-linear factor in $I_{RHEED}$ change is insignificant. Figure S4 shows the numerically simulated $I_{RHEED}$ variations for the above-described QW's growth regimes with different contributions of the two main factors determining the $I_{RHEED}$ behavior.

Figure 1e schematically illustrates a mixed (F-M&BCF) growth regime, which is more likely to be realized during the growth of subcritical GaN QWs under moderately Ga-enriched conditions on atomically smooth vicinal planes. This growth mode was first theoretically described for conventional III-V heterostructures in the early 1990s by Myers-Beaghton and Vvedensky[76] as well as by Fuenzalida.[77] These authors proposed a generalized BCF theory that takes into account not only the incorporation of diffusing adatoms into atomic steps but also the finite probability of their nucleation into 2D-nanoislands on vicinal planes. This theory was then developed by Ratsch *et al.*,[78] who considered the influence of stress on the kinetics of these processes. In general, according to these models, it can be assumed that at high values of the characteristic ratio $F/D$ (where $F$ is the growth flux, $D$ is the surface diffusion of adatoms) 2D-epitaxial growth of GaN is determined by the 2D-nucleation processes of 2D-nanoislands on vicinal planes and with an decrease in this ratio the contribution of step-flow (BCF) growth mechanism is enhanced.

Then, for QWs grown under highly metal-enriched conditions ($\phi_{Ga}/\phi_{N2*} = 2.1$ and $\phi_{Ga} > \phi_{N2*} + \phi_{Ga}^{d}$), a significant increase in the rate and amplitude of $I_{RHEED}$ linear decay is observed, as shown in the bottom graph in Figure 1b. This indicates not only the absence of influence of the second ("oscillating") factor on $I_{RHEED}$, but is also likely associated with a third factor – the appearance of excess metal (Ga) on the GaN surface, which can effectively additionally scatter the e-beam.[70] The nominal thickness of the adsorbed Ga layer ($d_{GaN}$) increases linearly with growth time (according to $d_{GaN}(t) \sim (\phi_{Ga} - \phi_{N2*} + \phi_{Ga}^{d}) \cdot t$ up the formation of a thermodynamically stable Ga BL in accordance with the mentioned above theory of Northrup and Neugebauer.[74]

The accumulation of excess metal is confirmed by Figure 1d, which shows the AFM image of a control 1.5-ML-thick GaN QW grown under highly metal-enriched conditions ($\phi_{Ga}/\phi_{N2*} = 2.1$) on one of the *T*-series structures. The total volume of the observed surface nanoclusters with typical thicknesses of 1−2 nm and lateral dimensions of about 15 nm, is approximately equal to the volume of 1.5 ML of continuous Ga layer formed during the growth of the 1.5 ML GaN QW at these conditions (Figure S5, Supporting Information). The nanoclusters appear after the completion of GaN growth with a rapid drop in temperature due to the dewetting phenomenon, which is a thermodynamically driven transition from a continuous Ga adlayer to a Ga nanocluster phase.[79] In contrast, AFM image of a GaN layer grown with the same thickness at a lower $\phi_{Ga}/\phi_{N2*} = 1.1$ shows no evidence of metal excess (Figure 1e). All structures from *T*-series also exhibit a droplet-free surface, as shown in Figure S6, Supporting Information.

The absence of droplets on the surfaces of all grown MQW-structures confirms that excess gallium periodically formed during the QW growth under metal-enriched conditions is desorbed during the subsequent growth of the AlN barrier layers due to the lower cohesive energy of Ga–N bond (2.20 eV) compared to Al–N (2.88 eV).[80] This assumption was confirmed by direct $I_{RHEED}$ measurements in our previous work.[52]

Based on the above results and assuming layer-by-layer filling of the upper and lower MLs in $GaN_m$/AlN QWs with a fractional of $1< m <2$ML, one can predict the formation of two different types of quantum-sized BL-thick GaN objects, depending on the prevailing growth mode. In the case of growth under nitrogen-rich conditions leading to the dominance of the F-M growth mechanism, 2D-QDs (also called quantum disks) will nucleate on AlN atomic terraces (Figure 1f). Under opposite metal-enriched conditions, the above-described effect of increasing the surface mobility of adatoms should lead to an increase in the role of the BCF- mechanism, which will lead to the formation of two-dimensional quantum ribbons (QR) adjacent to the edges of the steps and extending along them (Figure 1g).

## 3. Ex situ studies of structural properties ML-thick GaN/AlN multiple quantum wells

### 3.1. X-ray diffraction and X-ray reflection

The MQW structures were investigated also using measurements of X-ray diffraction (XRD) θ/2ω-scans of symmetric reflections (0002) (Figure S7, Supporting Information), as well as X-ray reflectometry (XRR) curves and small-angle scattering curves, shown in Figure 2b,c respectively. Table 1 compares the nominal period of the MQW structures ($T_{MQW}^{nom}$) and *m,n* values with their experimental counterparts $T_{MQW}^{XRD(XRR)}$, $m^{XRD(XRR)}$,$n^{XRD(XRR)}$ estimated from XRD(XRR) data, respectively, as it has been described in Refs.51,52. The summary Figure 2c shows that both X-ray methods estimate the MQW period and its relative change with increasing nominal QW thickness quite accurately and discrepancies between the nominal and simulated values are less than ~5%. However, the accuracy of determining the absolute values of the QW's and barrier layer's thicknesses is significantly lower. A similar discrepancy was observed in our previous works,[51,52] where it was associated both with the fractional filling of GaN atomic layers and the difficulties of correctly modeling the ML interface region between GaN and AlN atomic planes in the ultimately thin QWs.[81,82] Nevertheless, despite the difficulties described above, in a series of MQW structures with variable $m$ = 0.75–2 ML and a nominal period increment of 0.25 ML, X-ray methods make it possible to control the latter value with a fairly high accuracy (~3%). A continuous increase in the fractional value of $m$ from 1 to 2 can be considered as a gradual transition from the growth of structures with predominantly ML-thick QWs to the growth of structures with a predominance of BL-thick QWs.

Furthermore, Table 1 presents the results of modeling the morphology characteristics of the MQW structures using XRR-method. First of all, we note the significantly smaller RMS roughness of the structures estimated by this method ($RMS^{XRR}$) compared to the RMS values directly measured using AFM images shown in Figure 1a ($RMS^{AFM}$). This discrepancy is typical for these methods and can be explained by their fundamental differences, which is amplified when studying surfaces with strong deviations of the height distribution from the Gaussian shape.[83] Figure S8a,b, Supporting Information shows the non-Gaussian height distribution measured by AFM for the surface of the *T-1.25* ML structure, which can most likely be attributed to the formation of rare but pronounced topographic inhomogeneities in the form of small (with a diameter of < 50 nm) protrusions. Since $RMS^{XRR}$ calculations using standard software only take into account the Gaussian component of the height distribution, this leads to ignoring these small features, and hence to an underestimation of this roughness degree compared to $RMS^{AFM}$ data. However, the low $RMS^{XRR}$ value generally confirms the atomic smoothness of the upper AlN terraces and GaN/AlN interfaces in the near-surface MQWs. Note that the decrease of deviation of the height distribution from the Gaussian shape, which is observed for the *T*-series structures with larger $m$, as shown in Figure S8c,d, Supporting Information, leads to a decrease in the difference between the RMS values estimated by the two methods.

The gradual coarsening of the structures in *T*-series with increasing $m$ is also confirmed by Figure 2b, showing the transformation of the omega peak shapes from the sharpest in the *T-1.0* ML structure (i.e. in the absence of diffuse X-ray scattering) to its complete disappearance in the *T-2.0* ML structure with a thickness slightly smaller than the critical thickness of 2.4 ML for the 2D/3D transition in GaN/AlN QWs (Figure S1, Supporting Information).

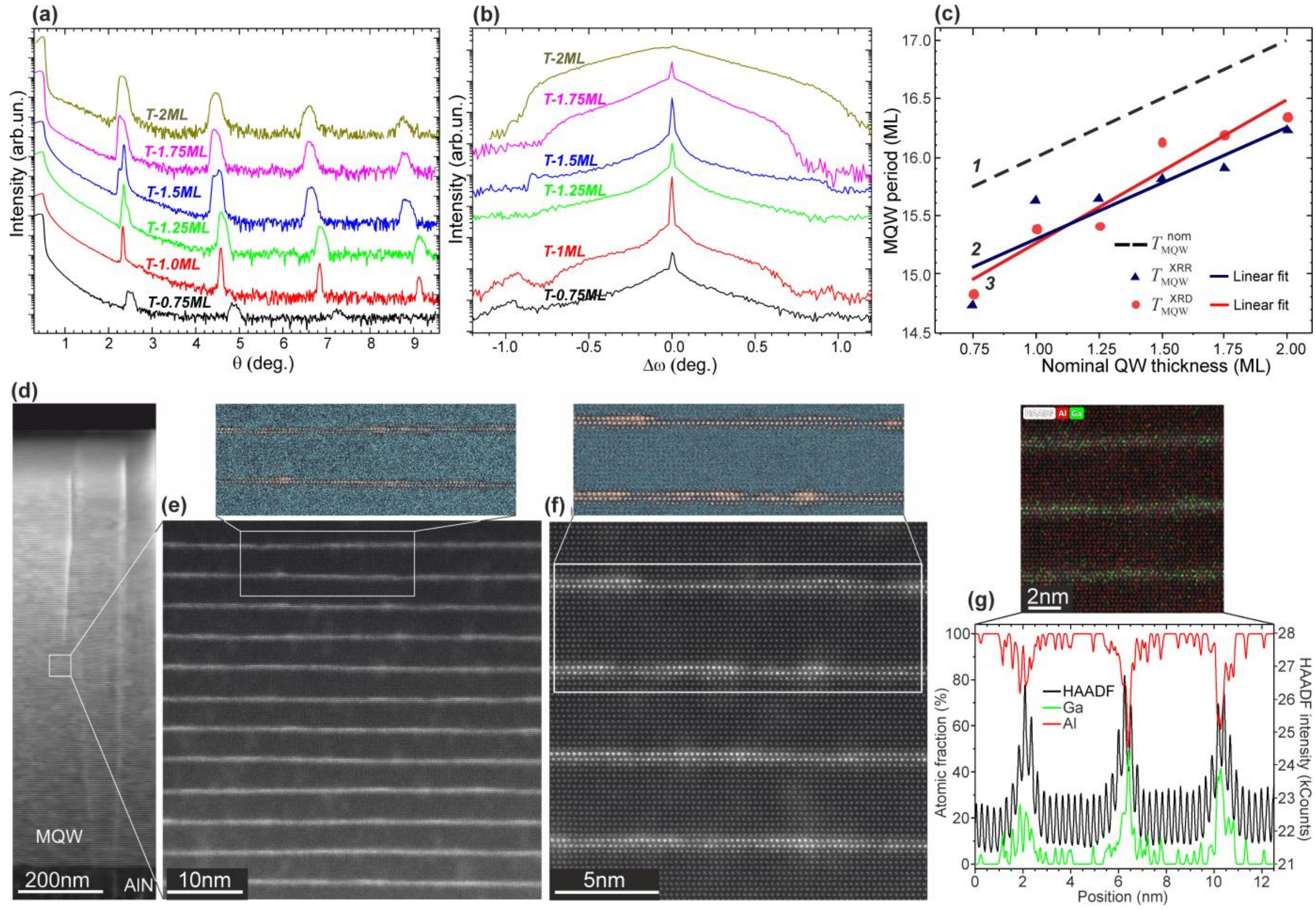

**Figure 2.** (a,b) θ- and ω-scans of the small-angle X-ray reflection curves for the *T*-series samples, respectively. (c) Dependences of the period of MQW structures on well thickness: (*1*) – $T_{MQW}^{nom}(m)$ nominal dependence, *(2)* and *(3)* – $T_{MQW}^{XRR}(m)$ and $T_{MQW}^{XRD}(m)$ dependences and their linear fits, respectively. (d-f) HAADF-STEM images of the MQW structure *F-1.6* with different magnifications. The insets in (e, f) show color enhanced images of the structure in the marked areas. (g) Combined HAADF-STEM and elemental mapping images of Ga and Al atoms obtained from energy dispersive X-ray spectroscopy data.

TABLE 1. Parameters of *T*-series structures simulated from XRD θ/2ω scans and XRR curves

| **Structure** | ***T-0.75ML*** | ***T-1.0ML*** | ***T-1.25ML*** | ***T-1.5ML*** | ***T-1.75ML*** | ***T-2.0ML*** |
|---|---|---|---|---|---|---|
| $T_{MQW}^{nom}$ (ML) | 15.75 | 16.0 | 16.25 | 16.5 | 16.75 | 17.0 |
| $m^{XRD}$ (ML) | 0.502 | 0.490 | 0.394 | 0.590 | 0.707 | 1.07 |
| $n^{XRD}$ (ML) | 14.337 | 14.904 | 15.020 | 15.542 | 15.490 | 15.273 |
| $T_{MQW}^{XRD}$ (ML) | 14.839 | 15.394 | 15.414 | 16.132 | 16.197 | 16.343 |
| $T_{MQW}^{XRR}$ (ML) | 14.739 | 15.635 | 15.639 | 15.807 | 15.904 | 16.225 |
| $RMS^{XRR}$(nm) | - | 0.04 | 0.06 | 0.13 | 0.22 | - |
| $RMS^{AFM}$(nm)* | 0.32 | 0.33 | 0.28 | 0.43 | 0.38 | 0.69 |
| $L^{lat}$ (nm) | - | 29.2 | 34.8 | 10.2 | 11.1 | 5 |
| $L^{vert}$ (nm) | - | - | - | - | 4 | - |

*- $RMS^{AFM}$ values estimated from the AFM images over an area of 1×1 μm² shown in Fig.1a.

X-ray modeling of the $T$-series structures also revealed a lateral correlation length ($L^{lat}$) of approximately 30 nm, i.e. approximately equal to the width of the atomic terraces measured by AFM ($\omega_T$ in Figure 1a) only for *T-1.0ML* and *T-1.25ML* structures. For structures with larger $m$, the $L^{lat}$ values decrease significantly (approximately to 5–10 nm), and further studies are needed to clarify the physical nature of this effect.

Modeling of the vertical correlation length ($L^{vert}$) showed that it corresponds to the nominal period of the MQW structure $T_{SL}^{nom}$~16 ML~4 nm only in one *T-1.75* ML structure with a weaker stepped relief. Therefore, the lack of vertical correlation in other structures can be tentatively explained by the random nature of the highly stepped morphology of these MQWs, which reduces the correlation of their height profiles when analyzing XRR curves.

**3.2. Scanning transmission electron microscopy**

Figure 2d–f shows the HAADF-STEM images of the 250×{$GaN_{1.5}/AlN_{15}$} MQW structure from $F$-series (*F-1.6*), in which the z-contrast makes the Ga atoms with higher atomic masses appear brighter compared to the Al atoms. The determined from these images values of the average MQW period correspond to the nominal value $T_{MQW}^{nom}$ = 41.235Å with an accuracy of less than 2%, which corresponds to the non-uniformity of the MBE fluxes in the PA MBE setup. Importantly, GaN QWs exhibit a 2D-topography with the same period throughout the 250×period structure with a total thickness of more than 1 μm. All the threading dislocations observed in the MQW structure (Figure 2d) originate only from the AlN/$c$-$Al_2O_3$ buffer layer.

When analyzing HAADF-STEM images, it is necessary to take into account the specimen thickness of about 50 nm needed for sufficient image intensity. On the other hand, in the case of the step-flow growth mechanism of the MQW structures, the planarity of the atomic layers is disrupted at distances exceeding the atomic terrace width ($> \omega_T$). This can lead to the observation of multiple atomic planes of Ga and to an incorrect estimate of the number of MLs in the GaN QWs. This effect is enhanced with increasing specimen thickness, and, therefore, can vary across a specimen in the case of thickness gradient, which typically occurs during specimen preparation for HAADF-STEM. Figure 2e shows a small-scale HAADF-STEM image of the MQW structure, in which the thickness of the GaN QWs appears to be greater in the lower part of the image, where specimen thickness is greater, compared to the upper part, in which this type of thickness error should be smaller.

HAADF-STEM is compatible with the existence of ML-height steps in the GaN, forming 2D-GaN regions of both ML and BL thicknesses in the AlN matrix, consistent with the step-flow and 2D-nucleation growth mechanisms. The typical lateral size of 2D-BL-GaN regions varies from several to tens of nanometers. At this stage of the study, it is difficult to assess the topology of the regions and the specific mechanism of their formation. This task is complicated by the complex topography of AlN barrier layers, consisting of hexagonal mounds with varying terrace widths and heights (Figure 1a), and the expected F-M&BCF mixed growth mode of GaN QWs (Figure 1e). Figure S9, Supporting Information presents the simulation of the HAADF-STEM images of $GaN_m/AlN$ ($m$ = 1−2 ML) QWs.

**4. Electronic properties of monolayer GaN/AlN quantum well structures**

The optical properties of III-N heterostructures are most often described theoretically either within the framework of density functional theory (DFT) or using a more resource-intensive quasiparticle approximation using the Green's function ($G_0W_0$).[84] An accurate description of the electronic structures of these heterostructures using the first method is problematic, since the electronic states in the conduction band are determined with an accuracy of up to a constant due to the discontinuity of the derivative of the total energy in the forbidden gap.[85] Therefore, $G_0W_0$ method was used for an accurate quantitative calculation of the electronic structure of the $GaN_m/AlN_n$ MQW structures ($m$ = 1,2 ML; $n$ = 11,12 ML). It is noteworthy that there are 24 atoms in the unit cells of $(GaN)_2/(AlN)_{10}$ and $(GaN)_1/(AlN)_{11}$, while the unit cell of equal periodical short period superlattice $(GaN)_8/(AlN)_8$ contains 32 atoms. The calculated effective band gap values ($E_g$*) were then used to construct a scissors operator to improve the accuracy of calculations of the electronic properties of the QW structures with different configurations, evaluated within less resource-intensive DFT method. Figure S10, Supplementary Information shows satisfactory agreement between the calculated and measured band gaps of several $GaN_m/AlN_n$ structures ($m$ = 1.5−18 ML, $n$ = 3−16 ML).[86]

First, the electronic structures of the $GaN_m/AlN_n$ SLs with various $m$ and $n$ values were calculated using $G_0W_0$ method. Figure 3a shows the band structure for $GaN_8/AlN_8$ symmetric SL, those $E_g$* = 3.2 eV is close to that of bulk GaN (3.4 eV). Figure 3b,c show that with a decrease in $m$ to 2 and 1, successive increase in $E_g$* by 1.2 and 0.5 eV is observed, respectively. According to these calculations and taking into account the systematic error of the $G_0W_0$ method (0.2 eV), the $E_g$* values in the $GaN_1/AlN_{11}$ and $GaN_2/AlN_{11}$ MQW structures are 4.945 and 4.434 eV respectively. These results are generally consistent with the theoretical and experimental data.[42,43,46,64,65] Calculations of electronic properties of the same MQW structures using the DFT method show the values of $E_g$*=3.719 and 3.174 eV, respectively. The similar values (0.511 and 0.548 eV) of the $E_g$* difference between ML- and BL-thick QW structures, determined by both methods allowed the less resource-intensive DFT method to be used in further calculations.

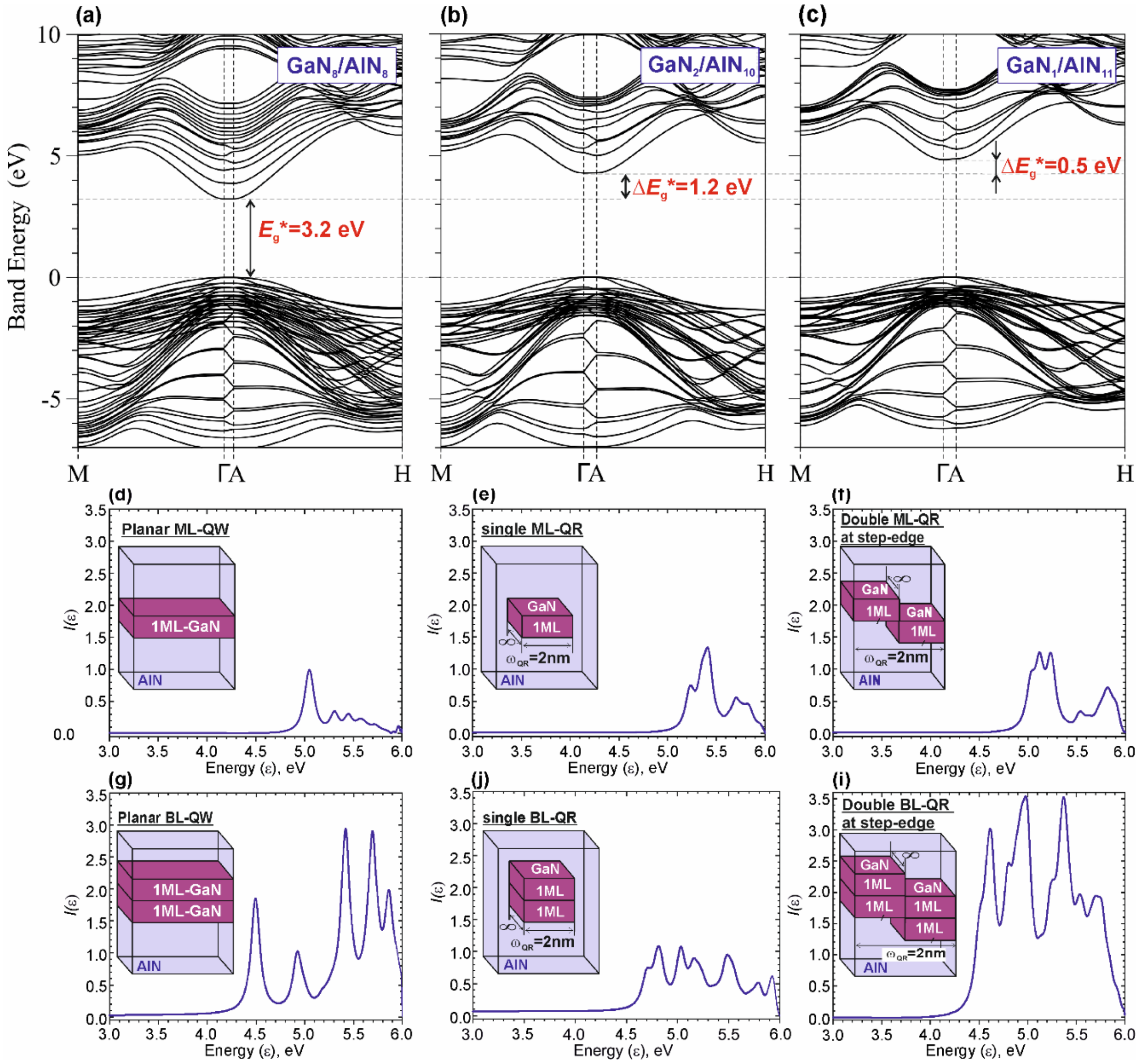

**Figure 3.** Electronic band structures of the symmetric $GaN_8/AlN_8$ SL (a), MQW structures $GaN_2/AlN_{10}$ (b), and $GaN_1/AlN_{11}$ (c), calculated using the $G_0W_0$ method. (d-i) Spectra of optical transitions in $GaN_m/AlN_{10}$ ($m$ = 1,2 ML) heterostructures with different configurations of GaN atomic layers shown (not to scale) in the insets of each figure. The graphs also show the calculated values of $E_g$* and intensities of optical transitions.

TABLE 2. Characteristics of quantum-dimensional GaN/AlN structures, determined from their spectra of optical transitions

| | Planar QWs | | Single QRs in AlN matrix | | Double QRs adjacent to step-edge | |
|---|---|---|---|---|---|---|
| | $E_g$*(eV) | $I$ | $E_g$*(eV) | $I$ | $E_g$*(eV) | $I$ |
| **1ML** | 4.929 | 0.99 | 5.120 | 1.35 | 4.912 | 1.27 |
| **2ML** | 4.381 | 1.86 | 4.577 | 1.09 | 4.387 | 3.02 |

Figure 3d–i show the spectra of optical transitions in the most interesting optical range of 3-6 eV, calculated by the DFT method using supercell approach for various configurations of GaN/AlN QWs with both ML and BL thicknesses. In addition to calculating the spectra of infinite planar QWs of different thicknesses (Figure 3d,g), we also calculated the spectra for so-called QRs with the same width of 2 nm (limited by the computational resources used) in one direction and an infinite size in the other direction. For each QR's thickness, two configurations were analyzed: single QRs in a homogeneous AlN matrix (Figure 3e,j) and double QRs adjacent at the step edges (Figure 3f,i). The $E_g$* values for these structures were determined from the coordinate of the intersection of the tangent to the low-energy edge of the spectrum with the energy axis. To quantitatively characterize the brightness of the emission from the structures, the intensities of optical transitions (*I(E)*) were calculated using the general equation:[87]

$$I(E) = \sum\sum\int \frac{dk}{(2\pi)^3} F_{ij}(k)\delta\left(E_{ik} - E_{jk} - E\right), \qquad (1)$$

where $i$ and $j$ are indices of occupied bands, respectively, and momentum matrix elements $F_{ij}$ are calculated for a given direction $\hat{u}$ using equation:

$$F_{ij} = \left|\langle\psi_{ik}|\hat{u}r|\psi_{jk}\rangle\right| \qquad (2)$$

The $E_g$* and $I$ values determined by DFT-simulation for all structures shown in Figure 3d-i are summarized in Table 2. The data for the infinite planar ML- and BL-thick QWs, in addition to confirming a significant difference between their $E_g$* of approximately 0.5 eV, showed a significantly higher intensity of optical transitions (absorption and emission) for the latter with the lowest value of $E_g$* value.

This analysis reveals the quantum confinement effect in QRs of both thicknesses, namely, an increase in $E_g$* by approximately 200 meV compared to planar QWs. Regarding the $I(E)$ of the QRs, only a small increase is observed for single ML-thick QRs, while for BL-thick QRs, its magnitude is even smaller than for an infinite planar BL-QW. In contrast, a significant increase in $I(E)$ is observed in the cases of double QRs adjacent to the step edges, and this effect is especially pronounced for BL-thick QRs (Figure 3i). Since this QR's location is thermodynamically more favorable compared to freestanding ones, one can expect an important role of such adjacent QRs in the optical properties of GaN/AlN MQW heterostructures.

## 5. Optical Properties of fractional monolayer GaN/AlN quantum wells

Figure 4a,S11a, Supporting Information show the RT and 77 K PL spectra of the *T*-series structures, which demonstrate change in the spectral position of the PL peaks from 228 to 258 nm (5.45–4.81 eV) with a gradual increase in the $m$ from 0.75 to 2 ML in 0.25 ML increments. Approximately the same shift in the position of the PL peak, accompanied by an increase in the integrated PL intensity, is observed in the *F*-series structures with an increase in the $\phi_{Ga}/\phi_{N2*}$ ratio, which is shown in Figure 4c, S11b, Supporting Information.

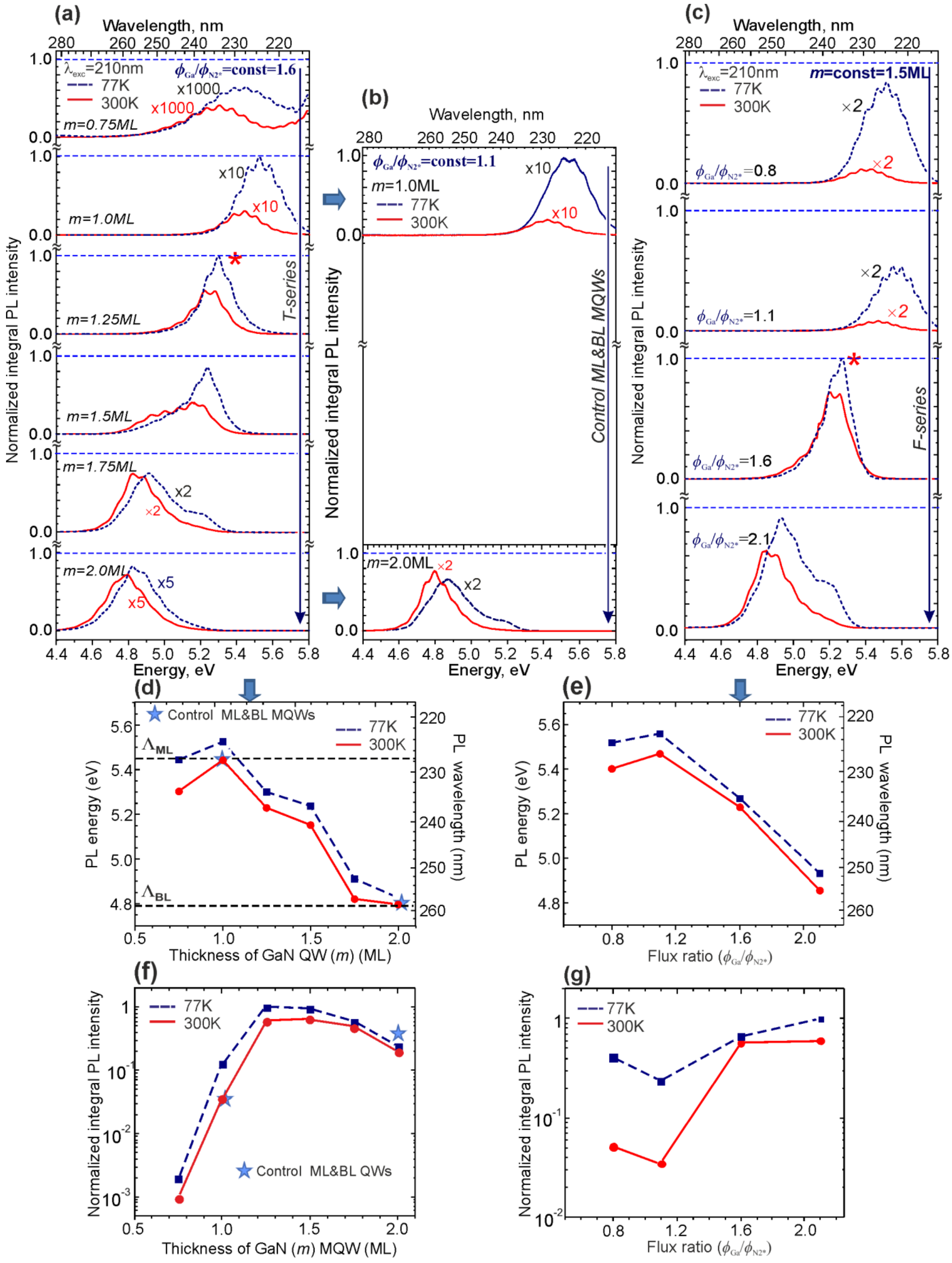

**Figure 4.** PL spectra (77&300 K) of $250\times\{GaN_m/AlN_{15}\}$ *T*-series MQW structures (a) and two control structures with $m = 1,2$ ML grown at $\phi_{Ga}/\phi_{N2*} = 1.1$ (b). (c) PL spectra (77&300K) of *F*-series MQW structures. All PL intensities are normalized relative to the maximum PL peak intensity observed at 77 K in the structure marked with a red star in (a). Dependencies of PL spectral position (77&300 K) on QW thickness in *T*-series (d) and $\phi_{Ga}/\phi_{N2*}$ in *F*-series (e). Dependencies of normalized PL intensities (77&300 K) on QW thickness in *T*-series (f) and $\phi_{Ga}/\phi_{N2*}$ in *F*-series (g).

The spectral positions and relative intensities of the main peaks in the CL spectra of both *T*- and *F*-series of MQW structures are similar to those in the PL spectra of the same structures. This conclusion is valid for the CL spectra excited using both high- and low-current electron e-guns, which are shown in Figure 5a and S12, Supporting Information.

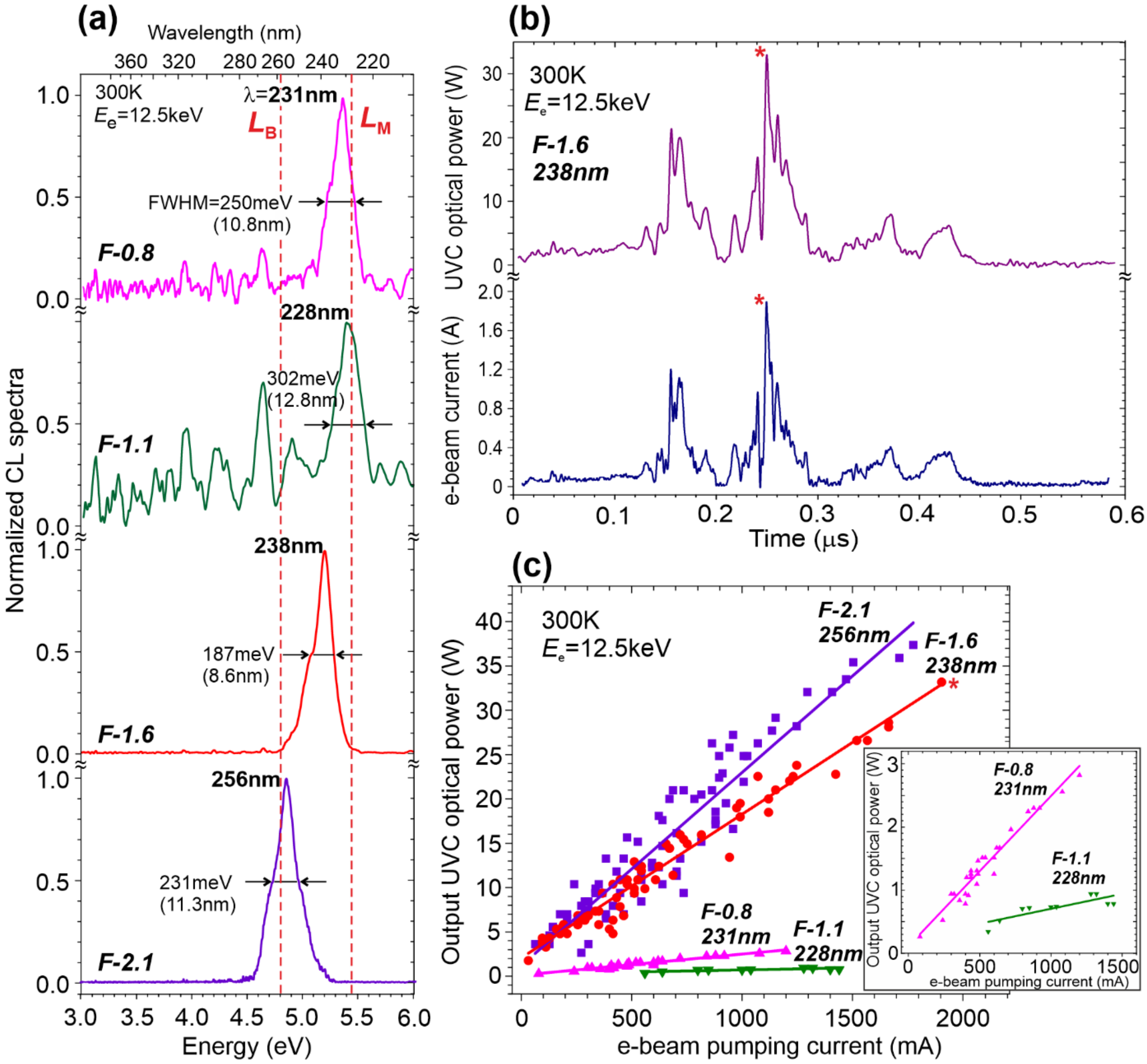

**Figure 5.** (a) CL spectra (300K) of the *F*-series structures excited by a high-current *e*-gun. (b) Oscillograms showing simultaneous changes in the output UVC-optical power from the *F-1.6* structure (upper curve) and the pumping pulsed e-current ($E_e$=12.5 keV, $I_e$ up to 1.9 A) (lower curve). (c) Current dependences of the peak optical output powers ($P(I_e)$) of the *F*-series structures, the inset shows magnified dependences for the *F-0.8, F-1.1*. The experimental data for the *F-1.6* structure, marked with an asterisk in (*b*) correspond to the same asterisk in the $P(I_e)$ dependence in (*c*).

The upper oscillogram in Figure 5b shows the temporal changes in the output UVC radiation emitted by the *F-1.6* structure when it is pumped by the high-current pulsed e-beam shown in the lower oscillogram.[88] Although each e-beam current pulse and the corresponding output UVC-radiation pulse have unique shapes, their integral characteristics are quite reproducible. Using these and other oscillograms for all *F*-series structures, the dependencies of the peak output UVC optical power (*P*) on the e-beam pump current ($I_e$) were plotted in Figure 5c,S13, Supporting Information. All MQW structures exhibit linear $P(I_e)$ curves, and the maximum peak power of $P_{\max} = 37$ W at a pump current of 1.8 A was observed in the *F-2.1* structure, which emits at the longest wavelength in this series, 256 nm.

Of more interest and practical significance are the results obtained for the *F-1.6* and *T-1.25* ML structures grown at the moderate Ga-rich conditions with $m = 1.25{-}1.5$ ML. The former structure not only emits a single narrow CL peak at 238 nm with a half-width of 187 meV (8.6 nm) but also ensures a $P_{\max} = 33$ W at $I_e = 1.9$ A, which is more than three times higher than our best result for this wavelength, demonstrated in our previous work.[52] Moreover, this structure exhibits an optical energy per pulse of up to 2.6 μJ, as shown in Figure S14, Supporting Information..

## 6. Discussion: optical properties of fractional-ML GaN/AlN MQWs vs their growth modes

The observed changes in the PL spectra in the *T*-series with varying nominal QW's thickness from 0.75 to 2 ML are plotted in Figure 4d,f. The red shift of approximately 600 meV observed in the first figure as well as for the CL spectra of the structures from this series (Figure S12) is consistent with the results of theoretical calculations of the electronic structures for ML and BL GaN/AlN QWs (Fig.3a−c) and therefore this range was denoted by the symbols $\Lambda_{ML}$ *and* $\Lambda_{BL}$.

Although MQW structures from the *T*-series with the thinnest QWs ($m \leq 1$ ML) exhibit UVC emission in the highly demanded 228−235 nm spectral range their intensities are extremely low. These results are consistent with the results of calculations optical transition intensities in 1-ML-thick quantum-sized objects, shown in Table 2. In addition, the relatively low ratio of PL intensities measured at high and low temperatures ($PL_{300}/PL_{77}$) indicates the low quantum efficiency of these MQW structures. This may be due to the absence of thickness and composition fluctuations in these QWs, which typically increase the efficiency of radiative recombination in QWs by forming a potential relief that limits the transport of charge carrier to non-radiative recombination centers (threading dislocations). Moreover, the relatively low CL intensity of these structures may also be due to the low efficiency of capture of charge carriers diffusing into extremely thin QW, as demonstrated by us.[89]

Therefore, we focused on the fractional-ML MQWs structures with 1 ML < $m \leq 2$ ML since they exhibited much higher PL and CL intensities, as well as evidences of charge carrier localization (i.e. high values of $PL_{300}/PL_{77}$ ratio), as shown in Figure 4,5. We begin our analysis of these structures by discussing the filling of the first (lower) GaN ML in a QW. To do this, we compare the top PL spectra in Figure 4a,b measured in two MQW structures with the same QW thickness of 1ML, including *T-1.0* ML and the control structures, grown at significantly different $\phi_{Ga}/\phi_{N2*}$ values of 1.6 and 1.1, respectively. The very similar shape, intensities, and spectral positions of these PL spectra mean their identical optical and, therefore, structural properties, despite significantly different prevailing growth mechanisms (BCF or F-M) realizing during their growths. The same conclusion follows from a comparison of the bottom PL spectra in Figure 4a,b, measured for another pair of MQW structures with the same an BL-thick QWs, but also grown at different $\phi_{Ga}/\phi_{N2*} = 1.6$ and 1.1,respectively. Consequently, for all structures in the *T*- and *F*- series, one can assume a layer-by-layer filling of the first and second MLs in GaN QW. Moreover, these results indicates that all differences in the PL and CL spectra observed in Figure 4a,c,5a are related with the filling of the second (partially filled) ML only.

During the growth of the upper half-ML in the *F*-series at small $\phi_{Ga}/\phi_{N2*} = 0.8–1.1$, the dominance of the F-M growth mechanism described in Section 2, should result in the nucleation of individual BL-thick 2D-QDs. These structures not only emit in the demanded spectral range of about 230 nm, but also have a significantly higher intensity emission compared to the emission from the described above *T*-series structures with $m \leq 1$ ML.

The further significant increase in the UVC-emission intensity by more than an order of magnitude is achieved in the PL and CL spectra of *F*-series structures with the same 0.5 ML thickness of the upper ML, but grown under Ga-enriched conditions ($\phi_{Ga}/\phi_{N2*} \geq 1.6$). According to the results of Section 2, the growth of these QRs is largely controlled by the BCF growth mechanism, resulting in the formation of BL QR-like objects near the step edges. Therefore, the observed increase in the PL and CL intensities can be presumably explained by appearance of these objects. This assumption is confirmed by a comparative analysis of the PL spectra of the *F-1.6* and *T-1.25*ML structures (Figure 4a,c). In both structures grown under nearly identical Ga-enriched conditions, a high contribution from the BCF growth mechanism and, consequently, a high probability of QR formation emitting near the same wavelength of 240 nm can be expected. Assuming a layer-by-layer BCF growth mode, the upper limit of the QR width ($w_{QR}$) in the *T-1.25*ML structure can be estimated as 1/4 $w_T$ (~10 nm), as shown in Figure S15, Supporting Information. In reality, this value may be smaller due to the parasitic

F-M growth mechanism, leading to the formation of GaN 2D-QDs instead of QR expansion (see Figure 1g,S9, Supporting Information). QRs are also likely to have a more complex non-rectangular topology due to the anisotropy of the growth rate in different crystallographic directions.[90]

Further increases in the $\phi_{Ga}/\phi_{N2*}$ in the *F-2.1* and $m$ in the *T-1.5* ML structures, leading to an increase in the width of QR, although causing to a relatively small increase in the CL intensities to maximum values (the PL intensities remain unchanged), are accompanied by a "red shift of the peaks to the spectral region of 250−260 nm). Moreover, we note that the *T*-series structures with the maximum filling of the upper ML ($m = 1.75$, 2 ML) even demonstrate some decrease in the PL and CL intensities. This indicates the existence of an optimal partial filling of the upper ML in the QWs, which ensures the maximum efficiency of radiative recombination at a minimum wavelength. The qualitative model developed above, of course, is overly simplified and future research is needed to optimize the growth of 2D-GaN objects in the fractional-ML GaN/AlN MQW structures.

## 7. Conclusion

In conclusion, we investigated the PA MBE growth of $250\times\{GaN_m/AlN_{15}\}/AlN/c\text{-}Al_2O_3$ structures with a nominal QW's thickness $m = (0.75–2) \pm 0.25$ ML under different growth conditions $\phi_{Ga}/\phi_{N2*} = 0.8−2.1$. RHEED studies revealed not only continuous 2D growth of these structures with stepped atomically smooth surfaces of AlN barrier layers, but also the growth of QWs according to 2D-nucleation (F-M) and/or step-flow (BCF) growth mechanisms. Based on experimental studies of these structures using AFM, XRD, XRR, HAADF-STEM, and PL measurements, as well as theoretical calculations using the $G_0W_0$ and DFT methods, we proposed a phenomenological model for the formation of fractional-ML $\{2D\text{-}GaN_m/AlN\}$ ($m \leq 2$) MQW structures with different sizes and topologies of quantum-sized active (light-emitting) regions. Depending on the nominal thickness and growth conditions, these ML- or BL-thick GaN regions can be controllably formed as 2D-QDs (or quantum disks) randomly distributed over the atomically smooth terraced AlN surfaces or QRs adjacent to and extending along the step edges. These MQW structures excited by e-beams with energies and currents up to 12.5 keV and 2 A, respectively, demonstrated the possibility of emitting UVC radiation with single peaks in the wavelength range from 228 to 256 nm and a maximum peak output optical power from 1 to 37 W, respectively, linearly increasing with the pump current.

**Experimental Section/Methods**

A. Experiment

The samples were fabricated by PA MBE (Compact 21T, Riber) on 2-inch $c$-$Al_2O_3$ substrates on which AlN nucleation and buffer layers with a total thickness of ~1.5 μm were grown using sequential migration-enhanced epitaxy and three-step metal-modulated epitaxy in the temperature range of 790−850°C, as described in Refs. 91,92. The growth of active regions with 250×{$GaN_m/AlN_n$} ($m$ = 0.75-2 ML, $n$ = 15ML) MQWs was described in detail in our previous paper.[51,52] Briefly, all structures were grown at the same substrate temperature of 690°C and $\phi_{N2*}$ of 0.55 ± 0.004 ML·s$^{-1}$, which was used to control the nominal thicknesses of $m$ and $n$, most of which were grown under metal-rich conditions. In one sample with QWs grown under nitrogen-rich conditions, a $\phi_{Ga}$ was used to control $m$. In all structures AlN barrier layers were grown under slightly Al-enriched conditions with a flux ratio of $\phi_{Al}/\phi_{N2*}$=1.1 and using BL-thick Al pre-deposited before the growth of each AlN. Also, before the growth of each QW, the AlN surface was exposed to an $\phi_{N2*}$ to consume excess Al. Design and designations of the two $T$- and $F$-series of the MQW structures are described in Section 2.1. In addition, two control 250×{$GaN_m/AlN_{15}$} MQW structures with integer $m$ = 1 and 2 ML were grown at $\phi_{Ga}/\phi_{N2*}$=1.1.

All growth stages were monitored *in situ* using reflection high-energy electron diffraction (RHEED), laser interferometry and multi-beam optical stress sensor. The study of surface of the MQW structures were carried out using atomic force and scanning electron microscopes (AFM and SEM, respectively). AFM study was performed on NTegra AURA (NT-MDT) scanning probe microscope using HA_NC (TipsNano) AFM probes. The MQW structures were characterized by X-ray diffraction (XRD) and small-angle X-ray reflectance (XRR) measurements using a D8DISCOVER Bruker AXS diffractometer with a rotating anode (CuKα1 radiation). XRD and XRR data were simulated using the Leptos software package version 7.04 from Bruker AXS. Omega curves of small-angle X-ray scattering were simulated within the Ming model included in the same software.[93] High-angle annular dark-field scanning transmission electron microscopy (HAADF-STEM) images were recorded using a spherical aberration-corrected FEI Titan Cubed Themis G2 300 transmission electron microscope (TEM) operated at 300 kV. The TEM samples were prepared by mechanical polishing and argon ion milling using Gatan PIPS™ Model 691.

The PL was excited by the radiation of a fourth-harmonic generator (Coherent) pumped by a Topol-1050-C-NC parametric femtosecond laser (Femtonika LLC) produced 80 MHz pulses with a duration of 150 fs. The excitation wavelength and average power density were 210nm and ~0.03 W/cm$^2$, respectively. The PL spectra, acquired at either 77 K or RT, were recorded using an Acton SP2500 spectrometer (Princeton Instruments) equipped with an 1800 grooves/mm diffraction grating and a liquid-nitrogen-cooled CCD camera (Pylon, Princeton Instruments).

The CL spectra were excited using both low- and high-current electron (e-) guns, which were recently reviewed in Ref.88. In the former, the e-beam with the energy $E_e$=10 keV was emitted by a standard incandescent e-gun with a tungsten cathode with the maximum value of the *cw*-electron current $I_e$=1 mA and the e-beam diameter on the sample surface $D_e$=1 mm. To generate high-current e-beams with $D_e$=4 mm, a pulsed e-gun based on a high-voltage surface discharge over a ferroelectric was used to provide the generation of e-current pulse packets with a total duration and frequency of ~0.5 μs and 12 Hz, respectively. The high-current e-beam had $E_e$~12.5 keV and the amplitudes of its current varied randomly during the discharge pulse up to 2 A. The e-beam current and CL spectra were registered using a miniature spectrograph with a concave diffraction grating, in which the signal was recorded on a CCD array with signal accumulation (see Section SII in Supporting Information).

B. Theory

The ab initio calculations of optical spectra of $GaN_m/AlN_{10}$ ($m$ = 1,2) heterostructures were carried out within local density approximation (LDA) to the density functional theory (DFT) using the pseudopotential method and the plane-wave basis set for wavefunctions as implemented in ABINIT software package.[94,95] The 2s2p, 3s3p, and 3d4s4p electronic states were considered as valence for Al,N and Ga atoms respectively. The expansion of wave functions on a plane-wave basis was controlled by completeness of the basis which determined by the cutoff energy $E_{cut}$=40 Ha. The reciprocal space sampling was performed on an 8×8×6 and 4×4×1 grid for bulk crystals and heterostructures correspondingly using the Monkhorst-Pack scheme.[96] The structural parameters were optimized for all structures until forces on atoms become less than $10^{-4}$ Ha·bohr$^{-1}$ and total stress below 0.1 GPa. The calculations of electronic structure were held with 150 unoccupied bands taken into account.

The calculations of electronic structure of limited set of ideal QW heterostructures $(GaN)_8/(AlN)_8$, $(GaN)_2/(AlN)_{10}$ and $(GaN)_1/(AlN)_{11}$ were additionally carried out using single shot $G_0W_0$ quasiparticle approximation.[97,98] The quasiparticle energies were calculated using

Kohn–Sham eigenstates and eigenvalues as an initial solution of non-interacting Hamiltonian. The inverse dielectric matrix was calculated using random phase approximation (RPA). The corrections to Kohn–Sham energies were calculated as [$\Sigma$-$E_{xc}$] operator diagonal matrix elements, where $\Sigma$=iGW—self-energy operator, $E_{xc}$ – exchange-correlation energy operator, G – Green function, and W – screening Coulomb interaction operator. The components of wavefunction with energies below 35 Ha for both exchange and correlation part were used to calculate $\Sigma$. The set of QW configurations structures with structural distortions were constructed using supercell approach.

**Supporting Information**

Supporting Information is available from the Wiley Online Library or from the author.

**Acknowledgements**

This work was supported by National Key R&D Program: Intergovernmental International Science and Technology Innovation Cooperation, Project №2022YFE0140100. The calculations were performed using facilities of the Ioffe computational resource center. The authors thank their colleagues at the Ioffe Institute for their assistance: Prof. V. Davydov, Dr. A. Smirnov, Dr. I. Eliseev, Dr. V. Kaibishev (optical measurements), Dr. S. Troshkov (SEM study), and Ms. D. Berezina (sample preparation), as well as Dr. D. Sviridov from the Lebedev Physical Institute (verification of AFM results), and Dr. A. Alekseev from SemiTEq (services for MBE setup). We also express our deep gratitude to Profs. H. Amano and M. Pristovsek (Nagoya University), and S. Ivanov (Ioffe Institute) for fruitful discussions and promotion of this research.

**Conflict of Interest**

The authors declare no conflict of interest.

**Author Contributions**

V.J. developed the initial concept of the Project and the layout of the paper; V.J., D.N., A.S. – designed, performed, and analyzed the epitaxial growth; E.R. conducted the theoretical studies, M.Y. measured and analyzed the XRD&XRR curves; P.A. carried out and analyzed the AFM study; E.E.,T.S.,A.T. performed the PL measurements and analyzed the optical properties; V.K., M.Z., N.G. performed and analyzed the CL measurements; T.W., carried out the HAADF-STEM study, V.J., T.S. co-wrote the manuscript with the participation of all authors, X.W., V.J. and A.T. supervised the project and provided its funding.

**Data Availability Statement**

The data that support the findings of this study are available from the corresponding author upon reasonable request.

# Supporting Information

## Fractional-monolayer 2D-GaN/AlN structures: growth kinetics and UVC-emitter application

Valentin Jmerik*, Dmitrii Nechaev, Evgenii Evropeitsev, Evgenii Roginskii, Maria Yagovkina, Prokhor Alekseev, Alexey Semenov, , Vladimir Kozlovsky, Mikhail Zverev, Nikita Gamov, Tao Wang, Xinqiang Wang, Tatiana Shubina, Alexey Toropov

## SI. Additional information

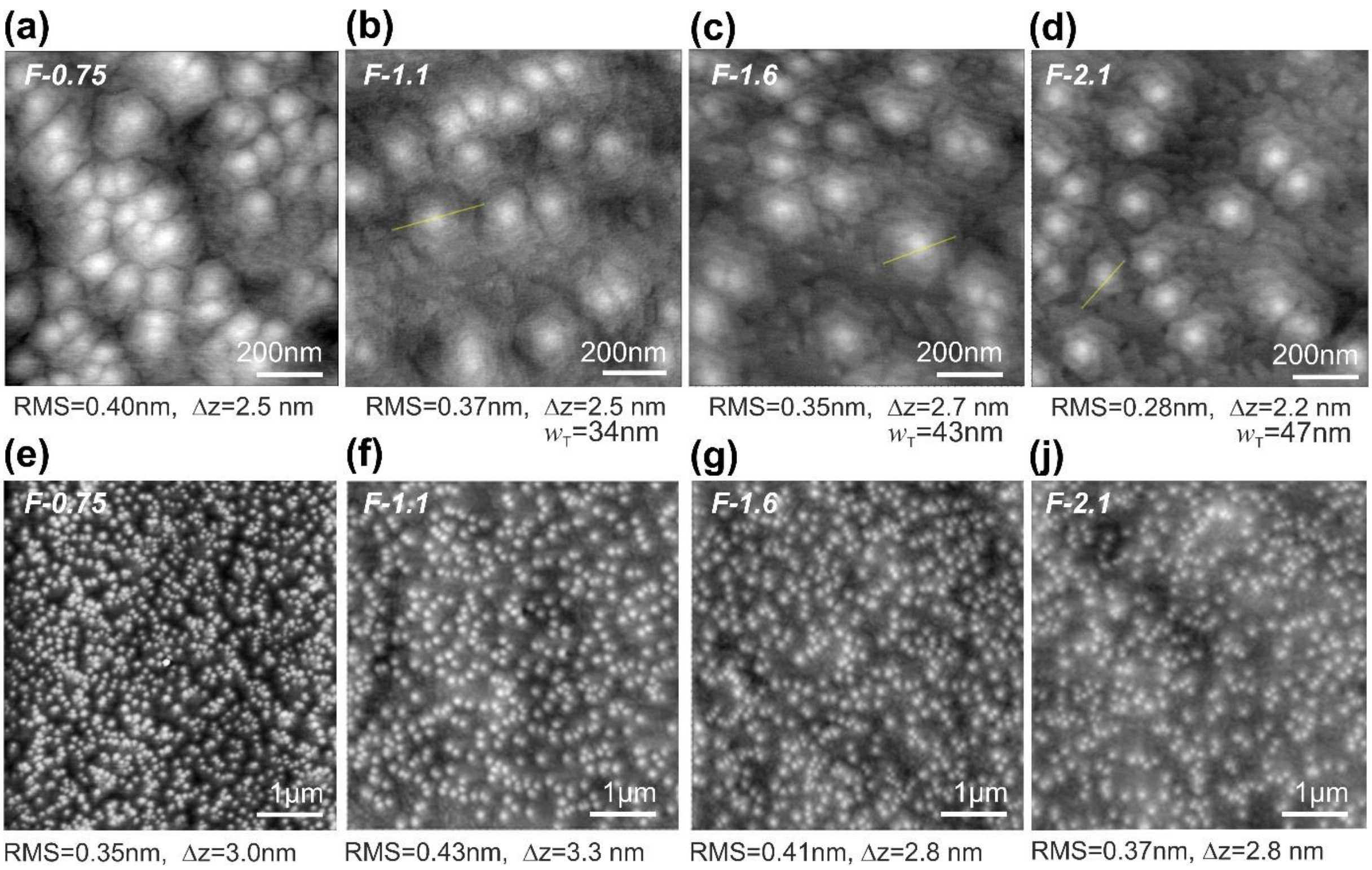

**Figure S1**. AFM images (1×1 μm$^2$) (a-d) and (5×5 μm$^2$) of the AlN top surfaces *F*-series 250×{$GaN_{1.5}/AlN_{16}$} MQWs structures grown at the different $\phi_{Ga}/\phi_{N2*}$: 0.8 (a), 1.1 (b), 1.6 (c), 2.1(d).

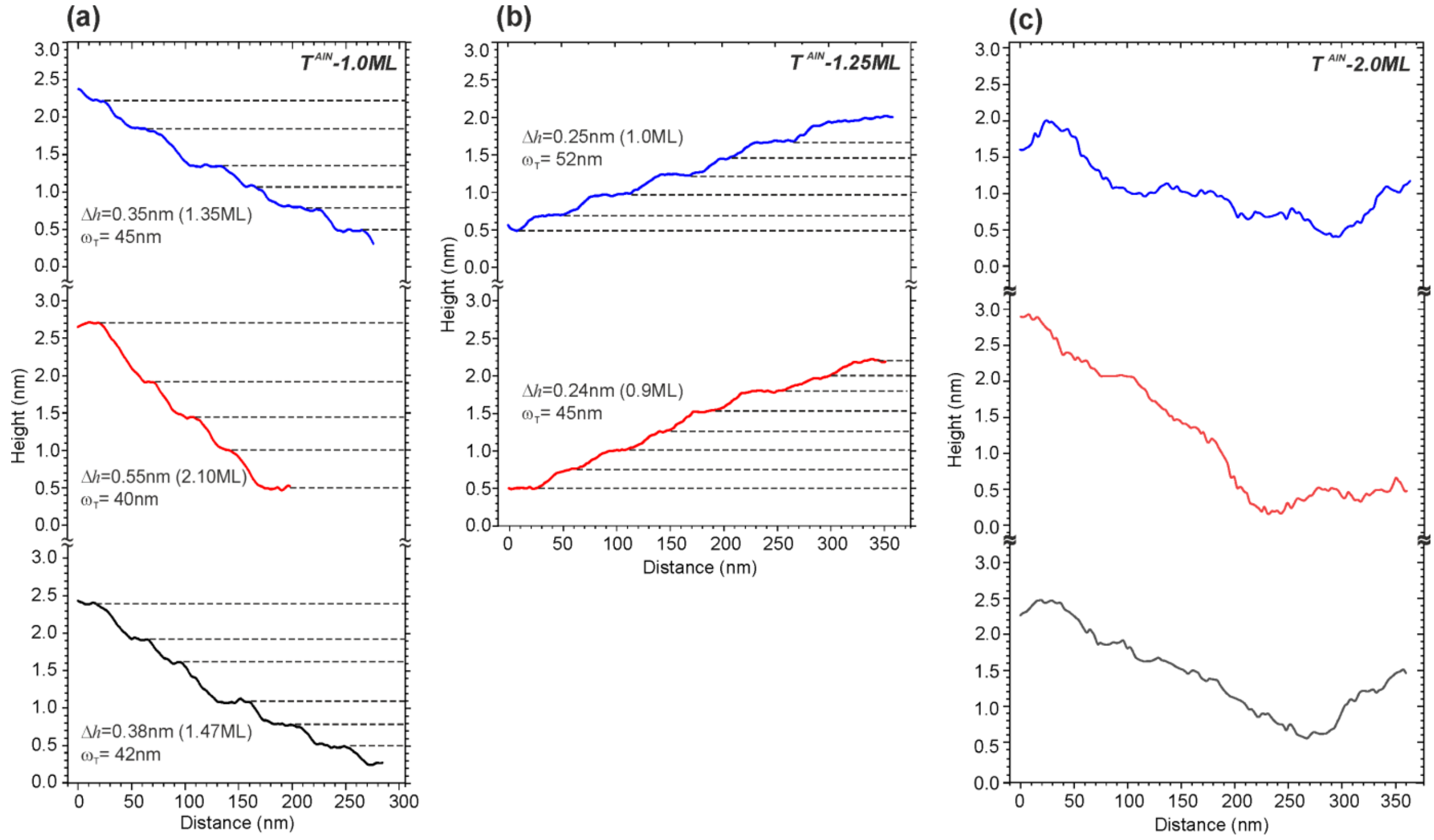

**Figure S2**. Height profiles measured at different positions in AFM images of the MQW structures *T*-1.0ML (a), *T* -1.25ML (b), and *T*-2.0ML (c). The average step heights ($\Delta h$) and terrace widths ($\omega_T$) are shown in (a) and (b) graphs.

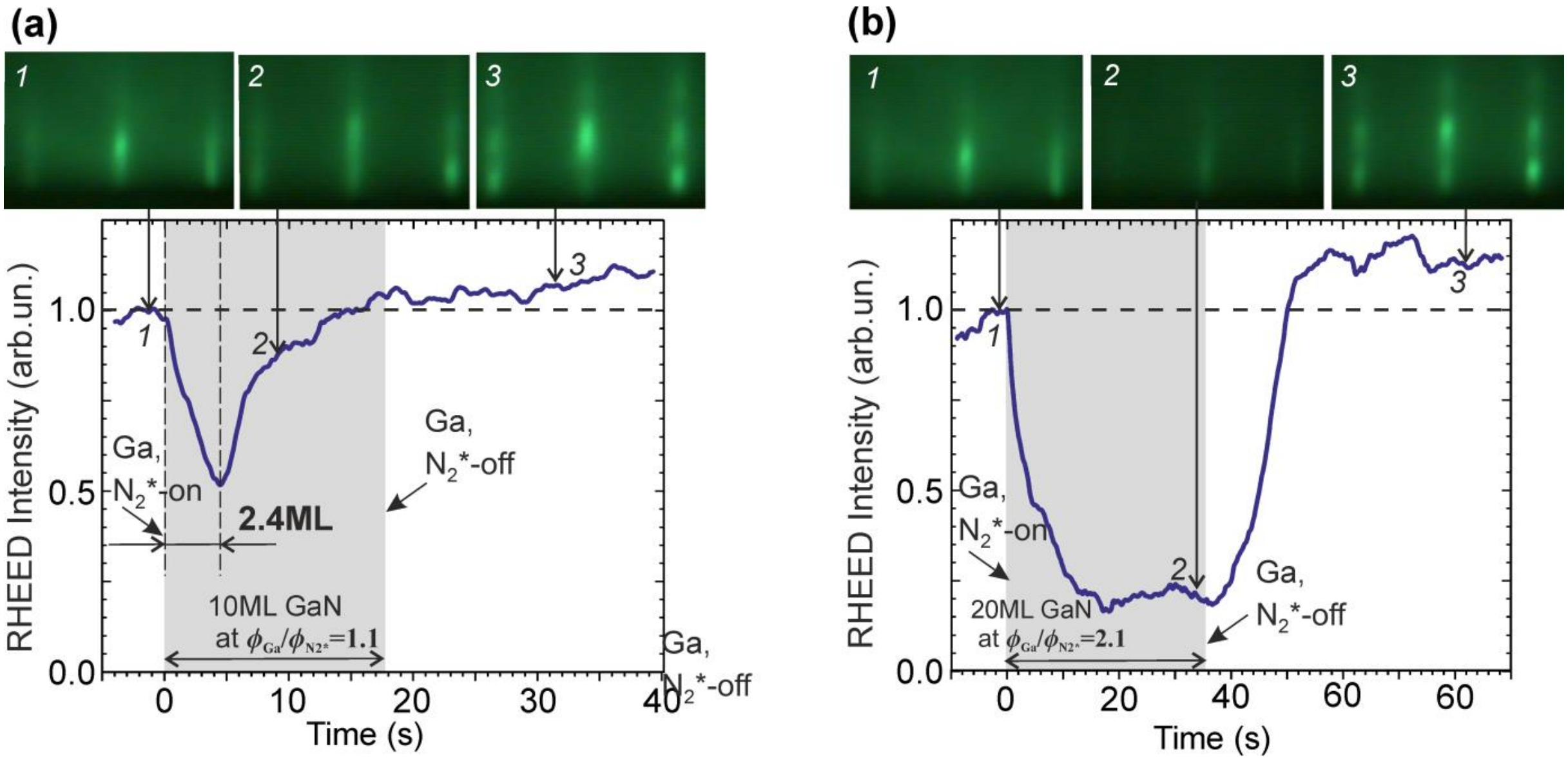

**Figure S3**. Time evolution of RHEED specular reflectance intensity during growth of GaN layers on an AlN with a thickness of 10ML at $\phi_{Ga}/\phi_{N2*}$ =1.1 (a) and with a thickness of 20 ML at $\phi_{Ga}/\phi_{N2*}$ =2.1 (b).

To investigate the GaN growth on the surface of AlN barrier layers in 250×{$GaN_{1.5}/AlN_{15}$}MQW structures, GaN test layers with different nominal thicknesses $t_{GaN}$ = 10 and 20 ML were grown at $\phi_{Ga}/\phi_{N2*}$ = 1.1 and 2.1, respectively. A full analysis of the critical thicknesses in these layers for the 2D-3D transition in the growth mode will be presented elsewhere. For this article, it is important that crtical thickness, corresponding to the transition from descending to ascending evolution of the RHEED intensity, exceeds 2 MLs.

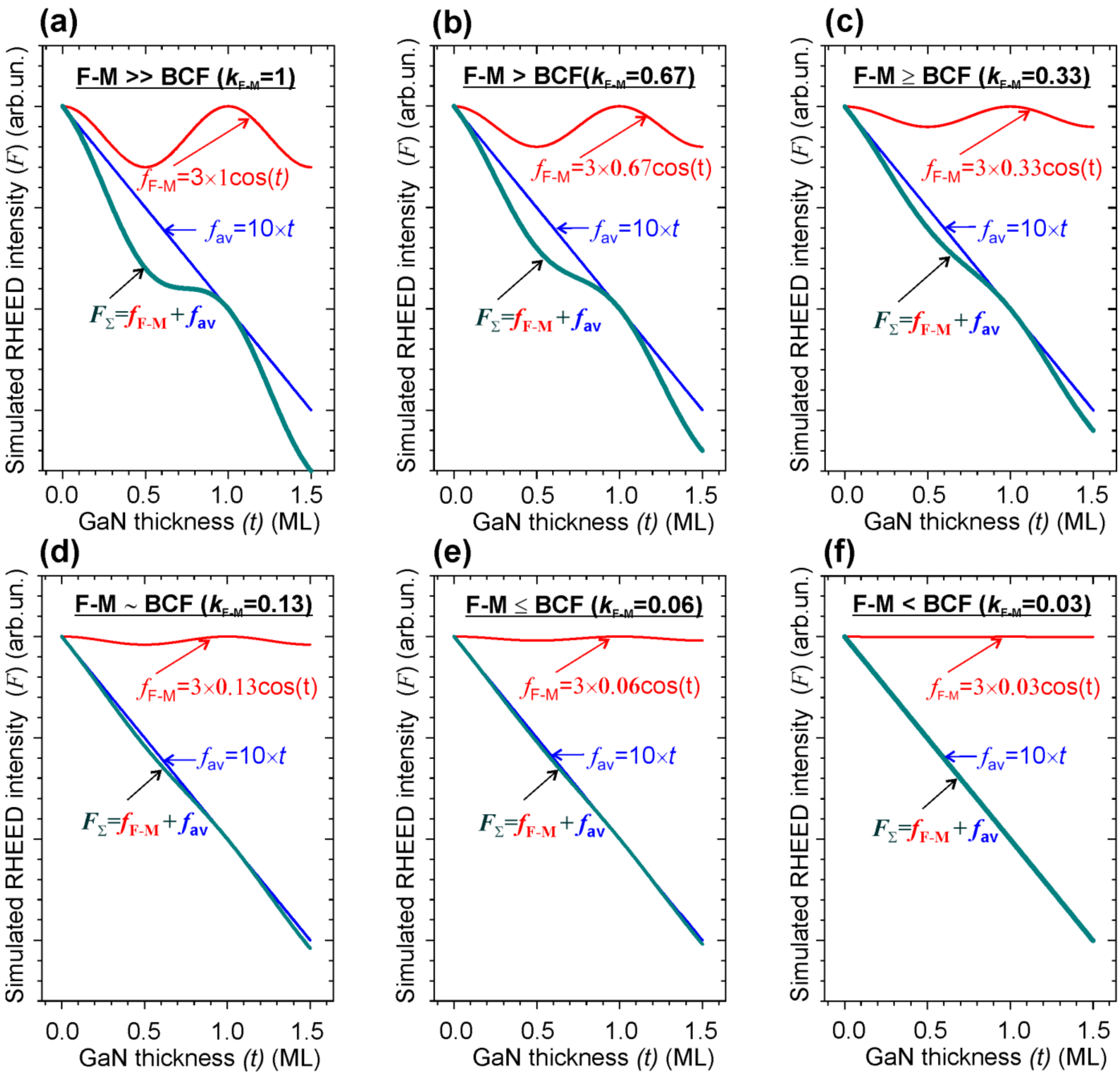

**Figure S4**. Simulation results of the temporal changes in the total RHEED intensity $I_{\Sigma}(t)=f_{av}(t)+f_{F\text{-}M}(t)$ (where $f_{av}(t)$, $f_{F\text{-}M}(t)$ are the linear and osillating factors described in the text below.

Modeling of changes in RHEED intensity was carried out under the assumption of the simultaneous influence of two factors on its change $I_{\Sigma}(t)=f_{av}(t)+f_{F\text{-}M}(t)$

- $f_{av}(t)$ - linear factor caused by the change in the average composition of the near-surface layer from AlN to GaN. In all simulated cases, the growth rate was assumed to be the same, and therefore the contribution of this factor was the same in all cases. The temporal behavior of this factor was described by a linear function $f_{av}(t)$=-$A$ t, where $A$

= 10 is a normalization coefficient determining the relative contribution of this factor to the overall change in IRHEED.

- $f_{F\text{-}M}(t)$ – is a factor of nonlinear (oscillating) $I_{RHEED}$ change associated with the gradual filling of the upper GaN layer and the formation of 2D-GaN islands in the case of 2D-nucleation (F-M) growth mechanism. In this simulation, we used the trigonometric approximation of this factor $f_{F\text{-}M}(t) = B\times k\cos(t)$, where $B$=3 - is the normalization coefficient determining the relative contribution of this factor to the overall $I_{RHEED}$ change, $k=v_{F\text{-}M}/(v_{F\text{-}M} + v_{BCF})$ – is the coefficient which characterizes the contribution of the F-M growth mechanism to the overall growth rate determining by both (F-M and BCF) mechanisms in the case of a mixed growth mode. In the Figure S4, the value of $k$ varied from 1 to 0.03, which corresponds to the dominant (F-M>>BCF) and insignificant (F-M<BCF) contributions of the F-M growth mechanism in the overall growth."

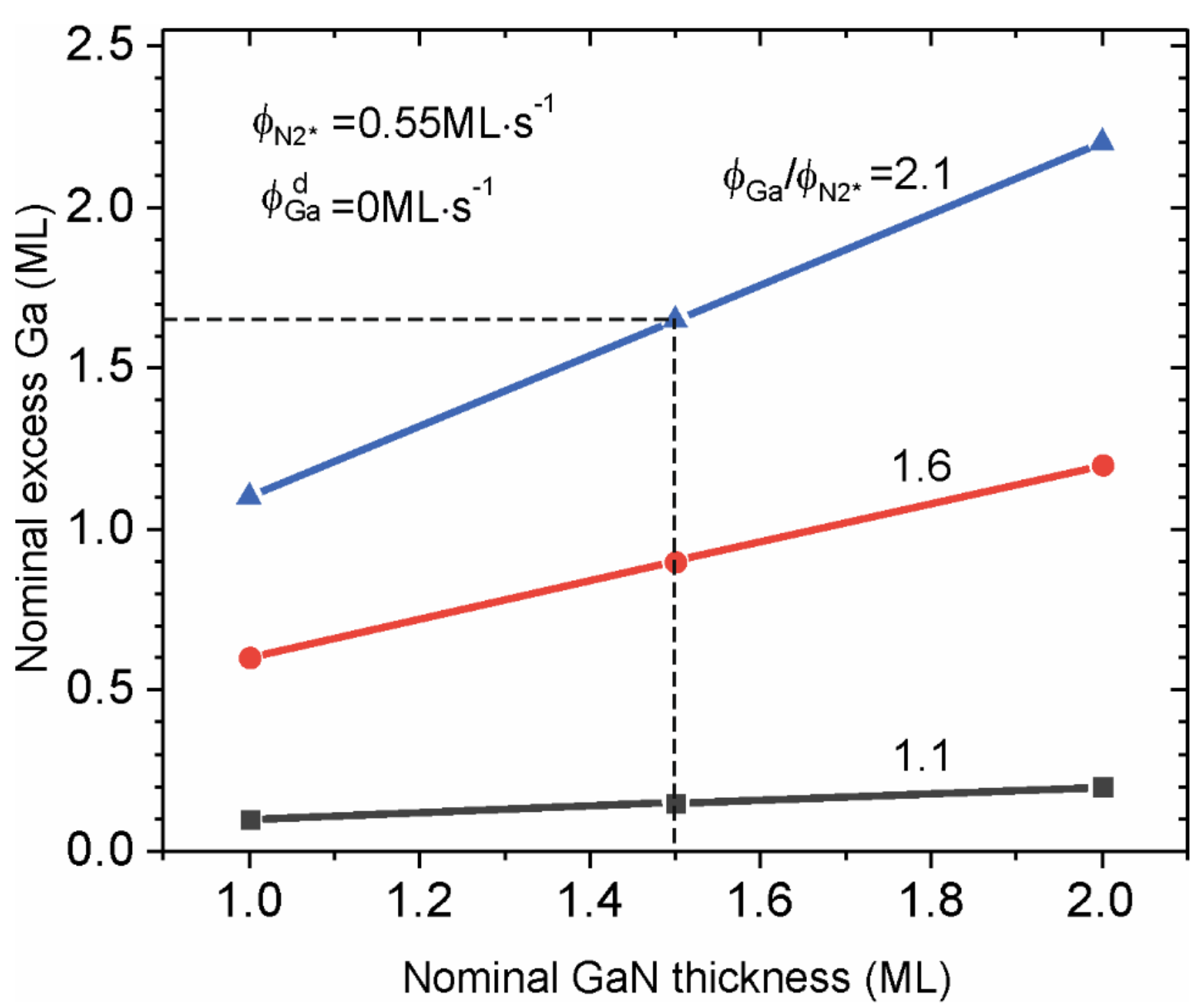

**Figure S5**. (a) Calculated dependences of nominal excess Ga adlayer thickness accumulated during the growth of GaN with a thickness from 1 to 2ML under different flux ratio $\phi_{Ga}/\phi_{N2*}$=1.1, 1.6, 2.1 and the same $\phi_{N2*}$=0.55 ML·s$^{-1}$, assuming negligible Ga-desorption ($\phi_{Ga}{}^{d}$=0).

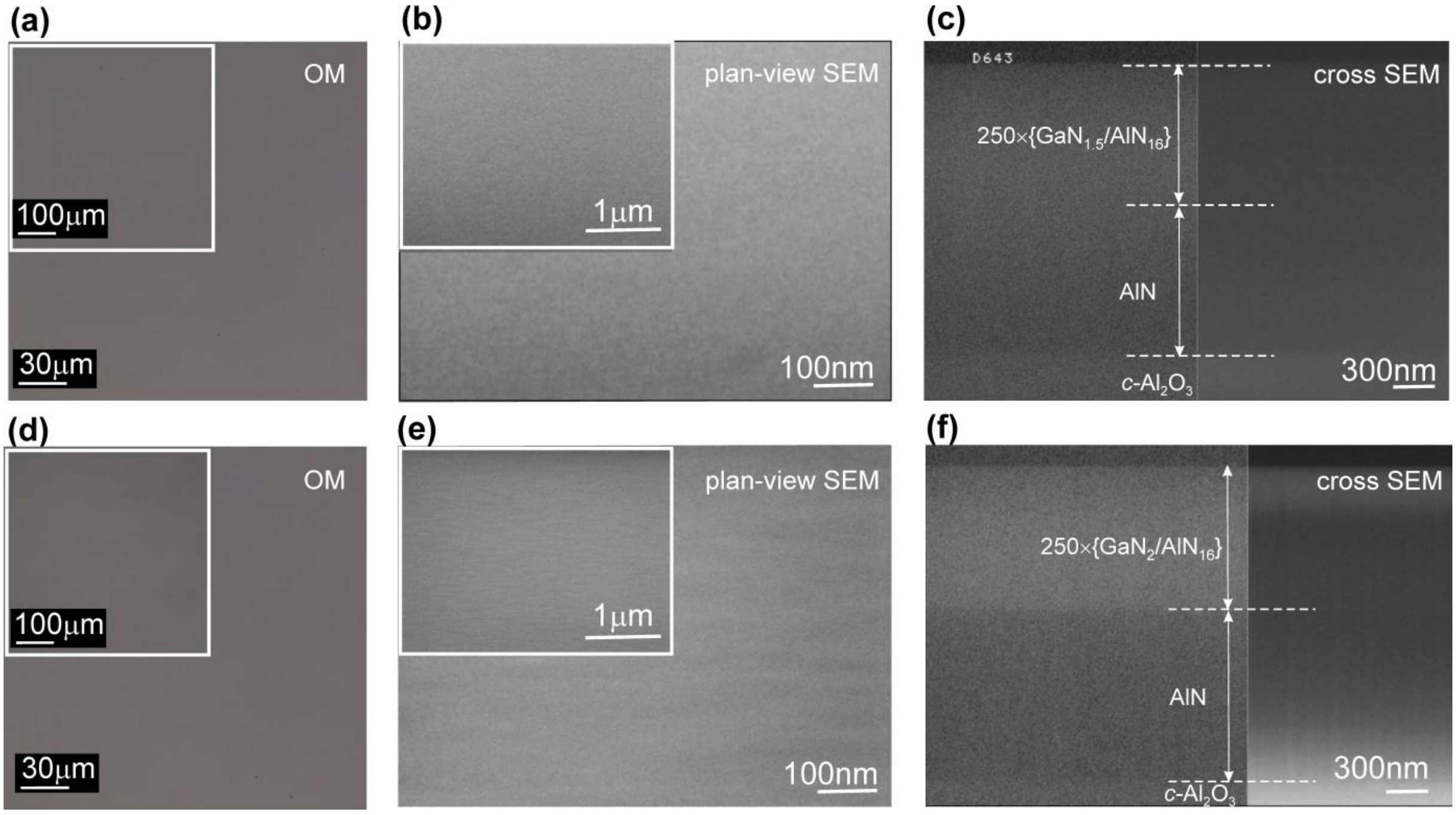

**Figure S6**. Optical microscope (a,d) and scanning electron microscope (SEM) images (b,c,e,f) of two structures *T-1.5 ML* (a-c) and *T-2.0 ML* (d-f). The cross-view SEM image (f) of 250×{$GaN_2/AlN_{15}$} MQWs on AlN/*c*-$Al_2O_3$ layer shows a contrast in Ga content betweeen these regions, whereas the analogous SEM image (c) of 250×{$GaN_{1.5}/AlN_{15}$} on the same AlN/*c*-$Al_2O_3$ does not show such a conrast.

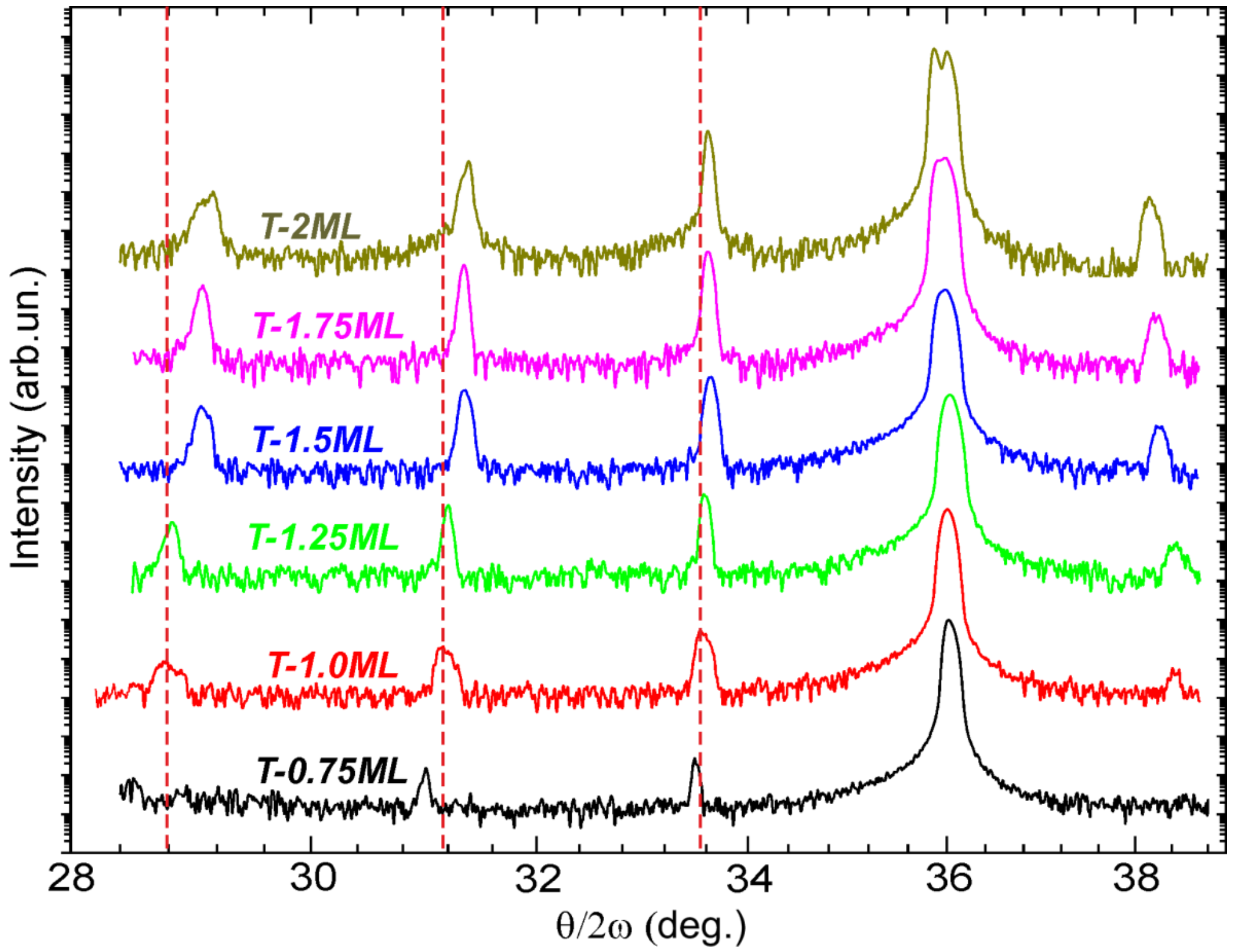

**Figure S7**. θ/2ω-scans of the X-ray diffraction curves of the symmetric reflection (0002) for the *T*-series 250×{$GaN_m/AlN_{16}$} MQW structures ($m$ = 0.75−2 ML).

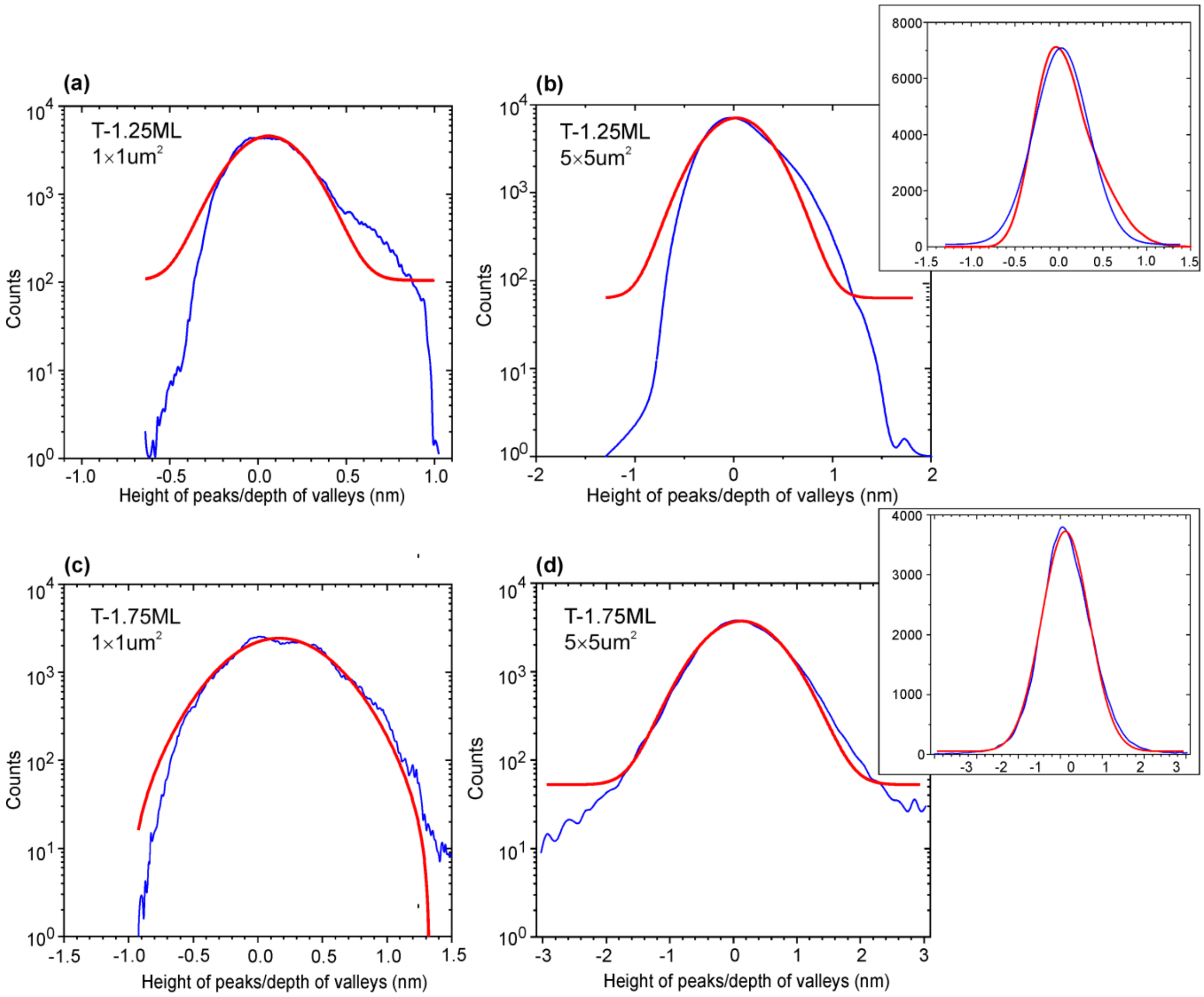

**Figure S8**. Distributions of height of peaks and depths of valleys for *T-1.25ML* (a,b) and *T-1.75ML* (c,d) structures measured using AFM over 1×1(a,c) and 5×5(b,d) μm$^2$ scan areas. The insets in Fig. (b) and (d) show the distributions in normal coordinates.

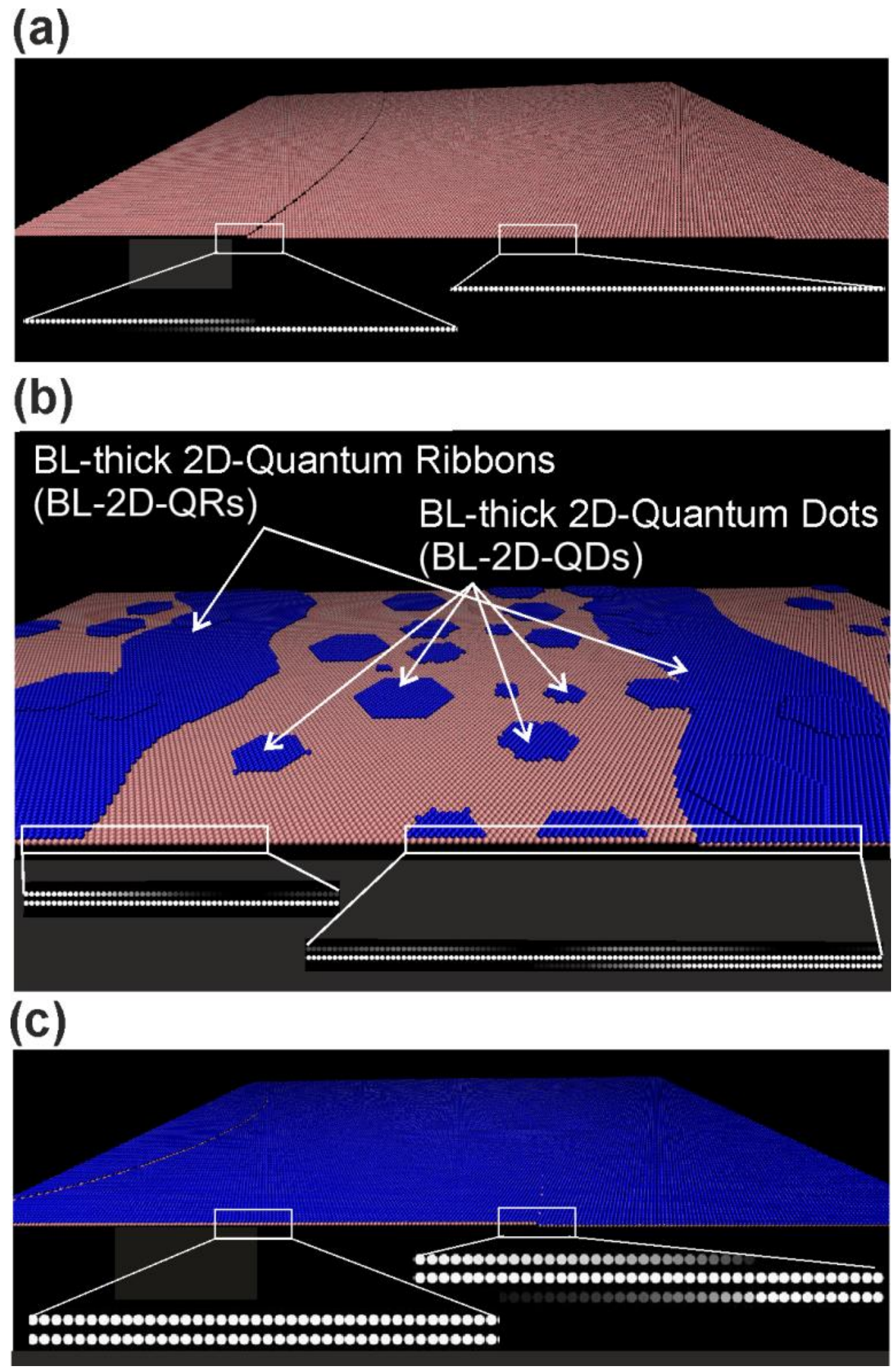

**Figure S9**. Schematic reperesentations of the $GaN_m/AlN$ QWs with various nominal thickness (*m*) and GaN relative content between QR ($n_{QR}$) and QD- ($n_{QD}$) phases in the upper layer: (a) *m*= 1 ML, (b) *m*=1.5 ML, $n_{QR}$=30 $n_{QD}$=20%; *m* = 2 ML(c) . These images were simulated using VMD 2.0 software, taking into account both the 2D-nucleation and step-flow growth mechanisms of fractional-ML GaN QWs. The bottom insets in each figure displays the corresponding thickness-averaged projections from the regions indicated by white rechtngles, which simulate cross-sectinal HAADF-STEM images of GaN QWs.

The accuracy of the $G_0W_0$ method was verified by comparing the calculated electronic structures of symmetric short-period $GaN_m/AlN_n$ $m$=$n$=3,4,6,8 ML superlattices, as well as asymmetric heterostructures with a constant QW thickness $m$=1.5ML and variable values $n$=2–16ML, grown, as described in Ref.81. Figure S10 shows satisfactory agreement (with a systematic error of ~-0.2 eV) between the effective band gap ($E_g$*) values calculated using the DFT method and the experimental values of this characteristic determined from the optical absorption spectra. Note that both the calculated and experimental values of $E_g$* begin to change only at barrier layer thicknesses less than 8 ML, which corresponds to the data of Sun *et al.*[S1], who calculated the critical thickness of the onset of overlap of wave functions of carriers in adjacent QWs to be equal to ~7–9 ML. Therefore, all calculations in this work were carried out for $m$=10 – 12ML, but their results should also be valid for MQW structures with larger AlN thicknesses, including $n$ = 15 ML.

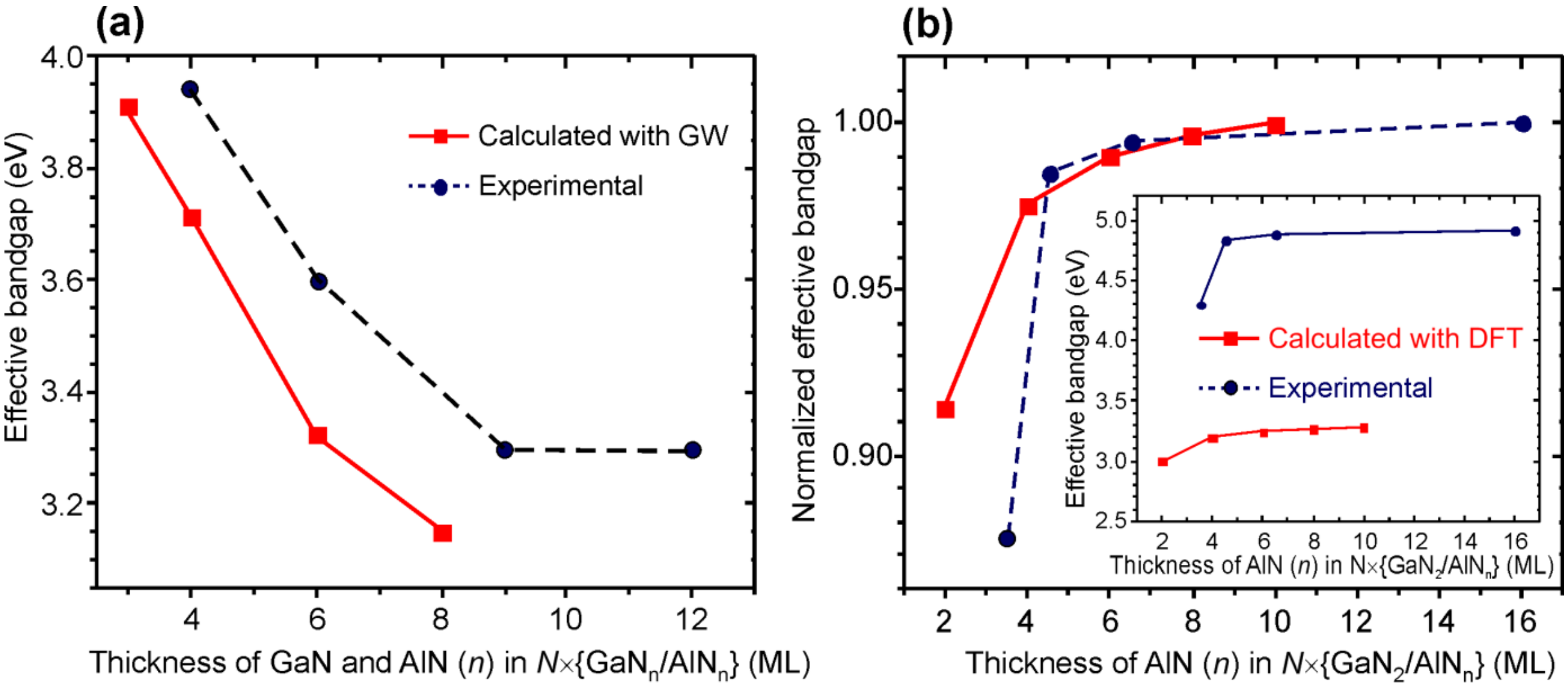

**Figure S10.** (a) The calculated by $G_0W_0$ method (red, solid) and experimental (blue, dashed) dependences of effective band gaps $E_g$* on AlN thickness (n) in N×{$GaN_n/AlN_n$} superlattices. (b) The calculated by DFT method (red, solid) and experimental (blue, dashed) dependences of effective band gaps $E_g$* on AlN thickness (n) in $N$×{$GaN_2/AlN_n$} heterostructures. In both experimental studies, the band gaps were determined from optical adsorption spectra.

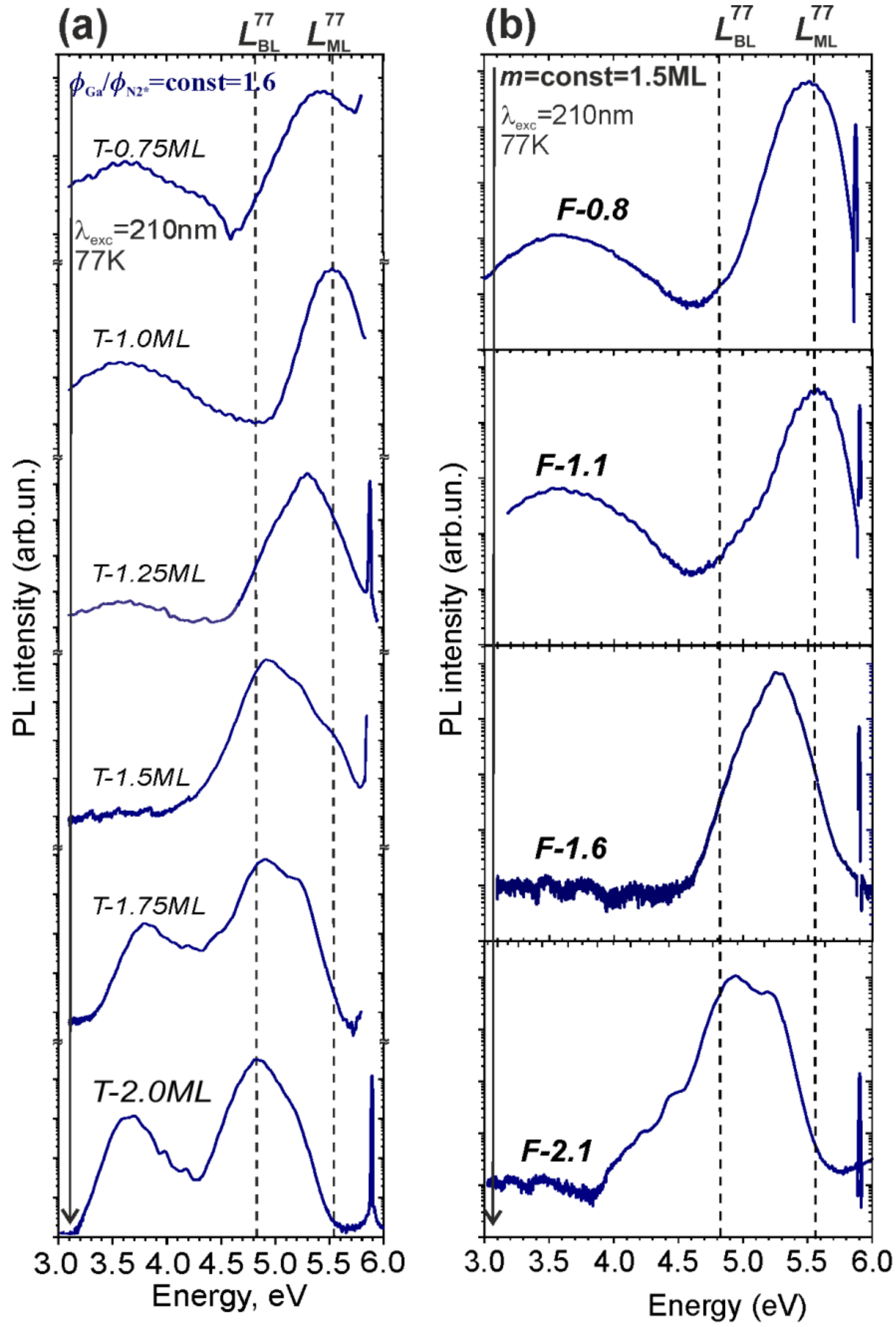

**Figure S11.** PL spectra (77K) of samples from the *T*-series (a) and *F*-series (b)., presented on a semi-logarthmic scale.

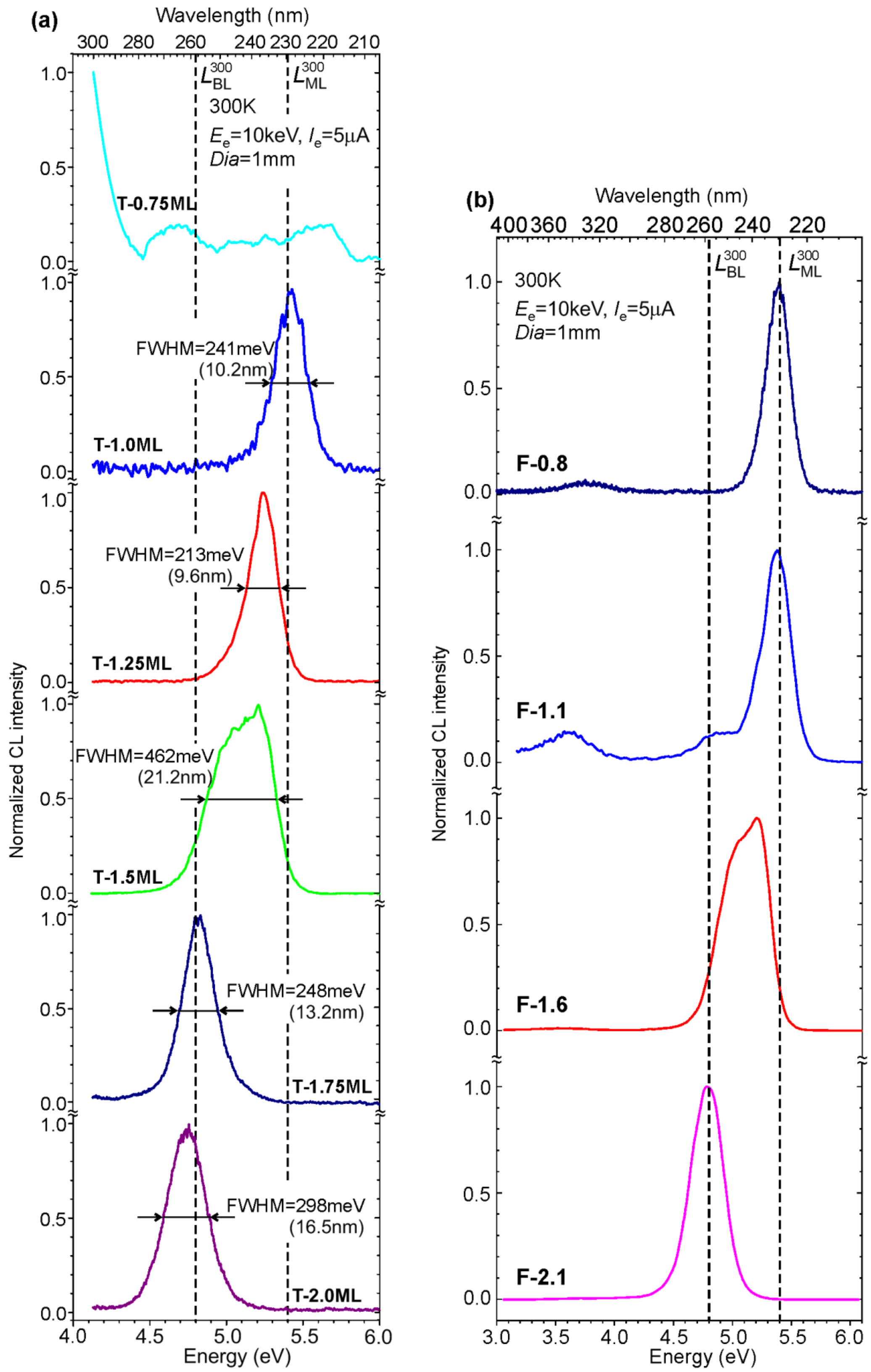

**Figure S12.** CL spectra (300K) of the *T*-series 250×{$GaN_m/AlN_{15}$} ($m$=0.75-2 ML) MQW structures (a) and *F*-series 250×{$GaN_{1.5}/AlN_{16}$} ($\phi_{Ga}/\phi_{N2*}$ = 0.8,1.1, 1.6, 2.1) (b), excited using a low-current e-gun with $I_e$=5 μA and $U_e$=10 keV.

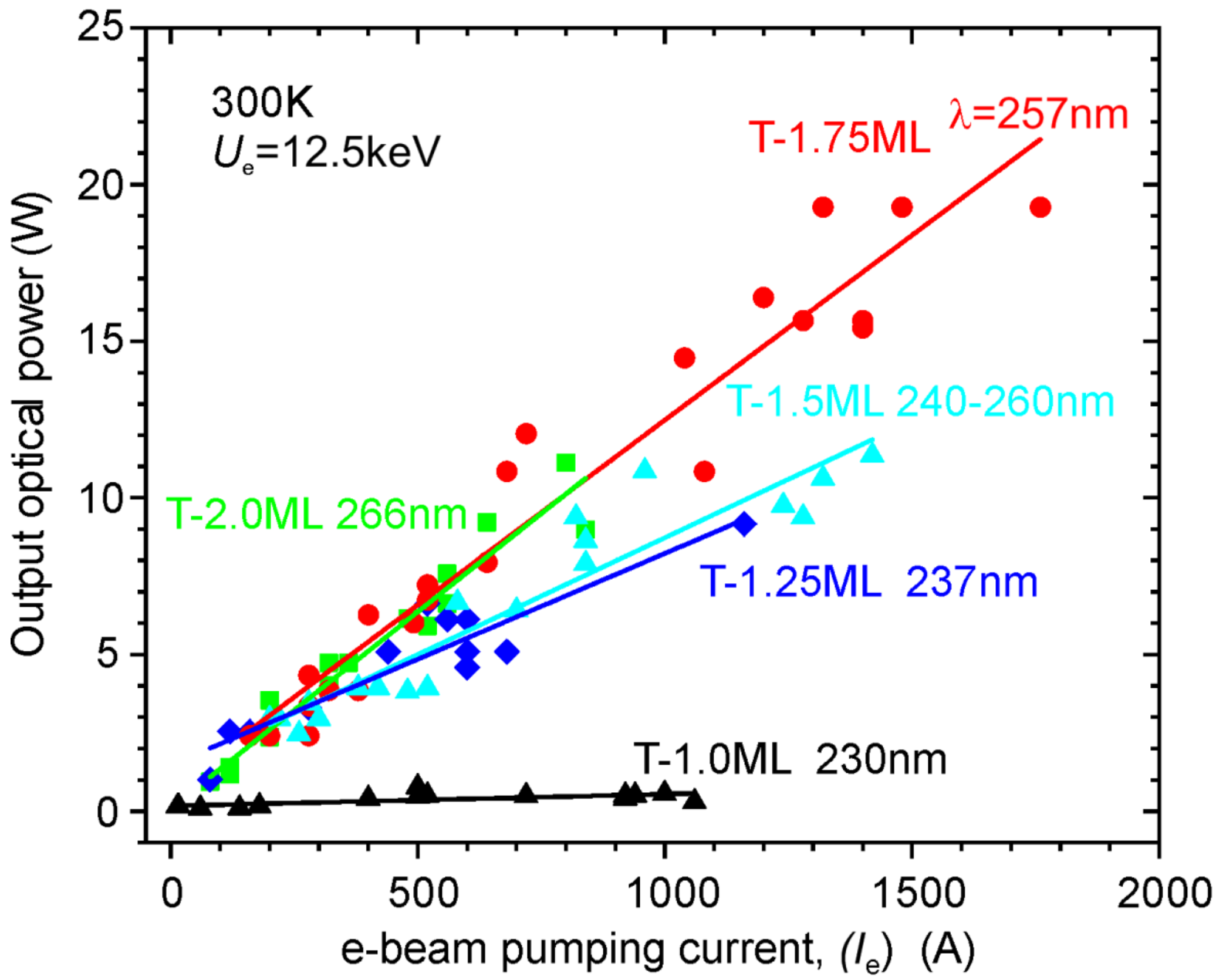

**Figure S13.** Current dependences of the peak optical output powers ($P(I_e)$) of the *T*-series structures: *T-1.0, T-1.25, T-1.5, T-1.75, T-2.0*, those low-current CL spectra are shown in Figure S11a.

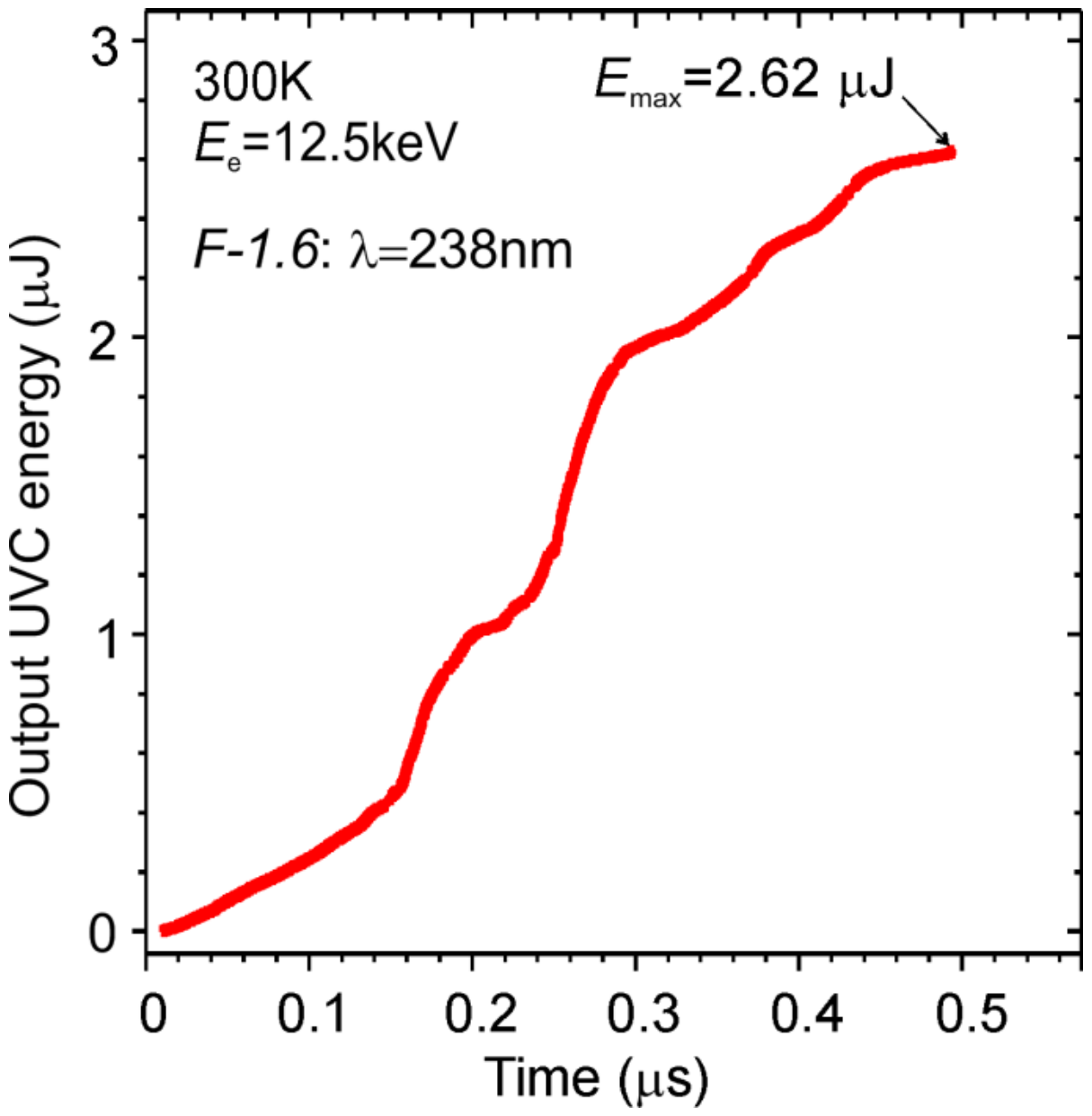

**Figure S14.** Time variation of the output optical energy of UVC radiation during one high current e-beam pumping pulse of the *F*-1.6 structure.

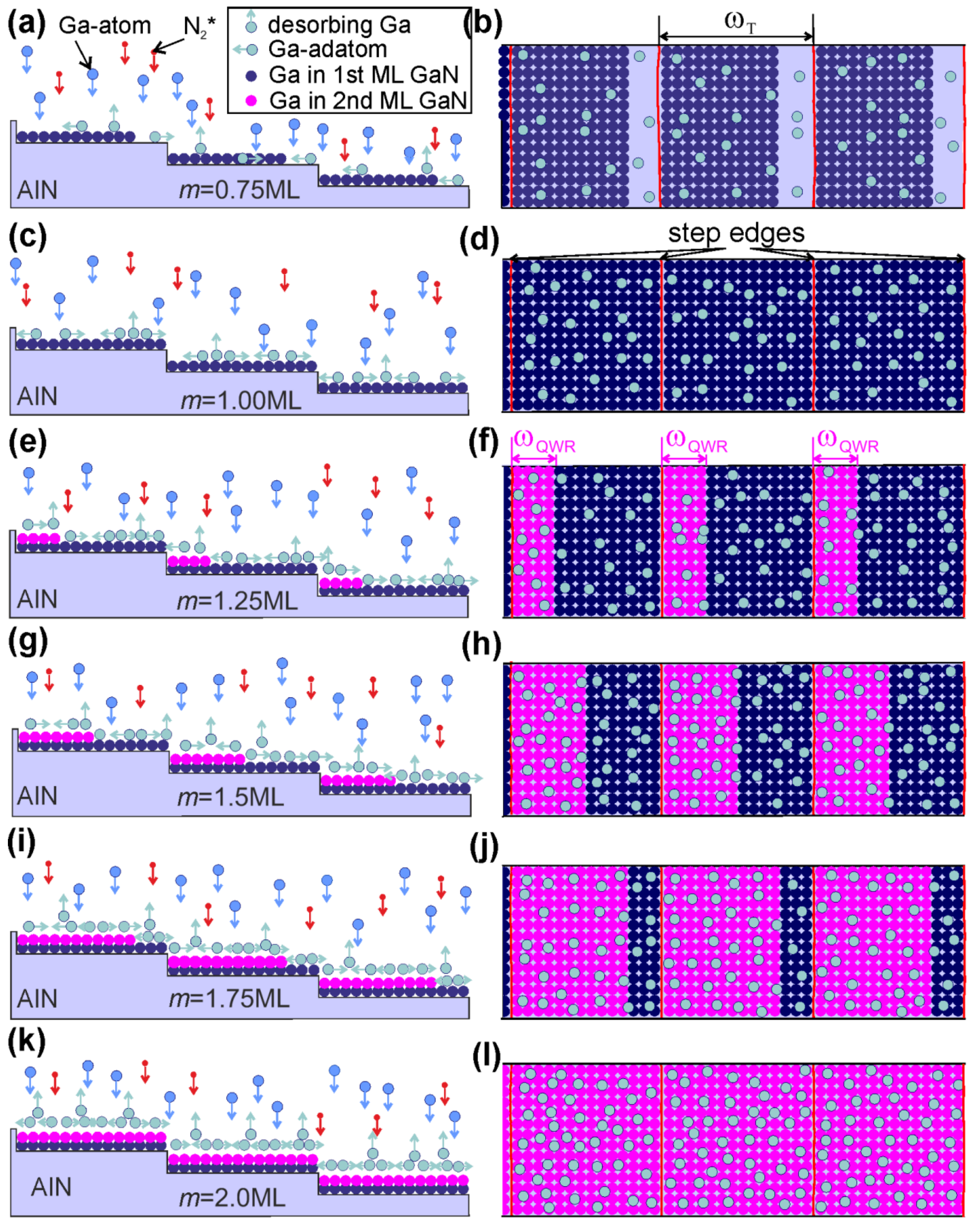

**Figure S15.** Scheme of the "ideal" layer by layer growth of GaN/AlN QW with $m = 0.75$-2 ML under Ga-enriched growth conditions in accordance with the step-flow (BCF) growth mechanism only (excluding the competitive 2D-nucleation (F-M) growth mechanism.

## SII. Electron-beam pumping of GaN/AlN heterostructures: design and technique for measuring the electron beam current, output optical power and optical spectra

### A. General scheme

Figure S16 shows a schematic of the setup used for e-beam pumping of GaN/AlN MQW heterostructures and detection of both the e-beam current and the output UVC radiation.

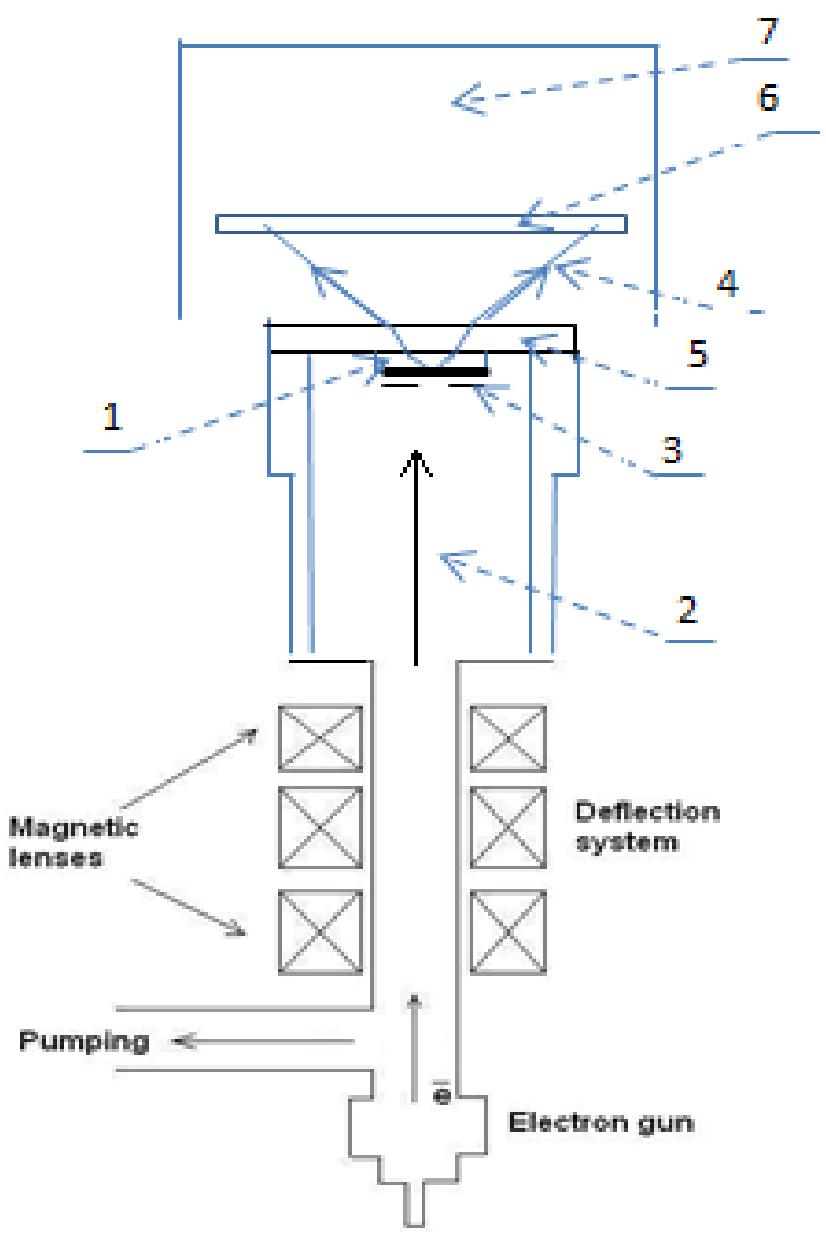

**Figure S16.** Scheme of the experimental setup, consisting of: *1* – sample with excited structure, *2* - electron beam, *3* - diaphragm in front of the sample, *4* - sample radiation, *5* - sapphire window of the camera, *6* - radiation detector photocathode, *7* - FEK-22 photodetector.

### B. Electron-beam gun with a plasma cathode

The setup above used an e-gun with a so-called plasma cathode, in which the plasma formed as a result of breakdown over the surface of a ferroelectric[S2]. The characteristics of such *e*-guns are detailed in Refs. S3–S6 and have been recently reviewed in Ref. 83. Figure S17 shows the specific design of a e-gun used in this work. The cathode of the electron gun was a ferroelectric plate 2–3 mm thick with metallized surfaces coated on the outside with a dielectric material. In one place, the plate is notched, and when a high voltage pulse from a pulse transformer was applied between the metallized surfaces, a breakdown occurred along the surface of the ferroelectric at the notch. The anode of the gun was a stainless-steel tube with polished surfaces, located at ground potential. The triggering frequency was determined by the repetition rate of the triggering pulses and was 1.5 or 12 Hz.

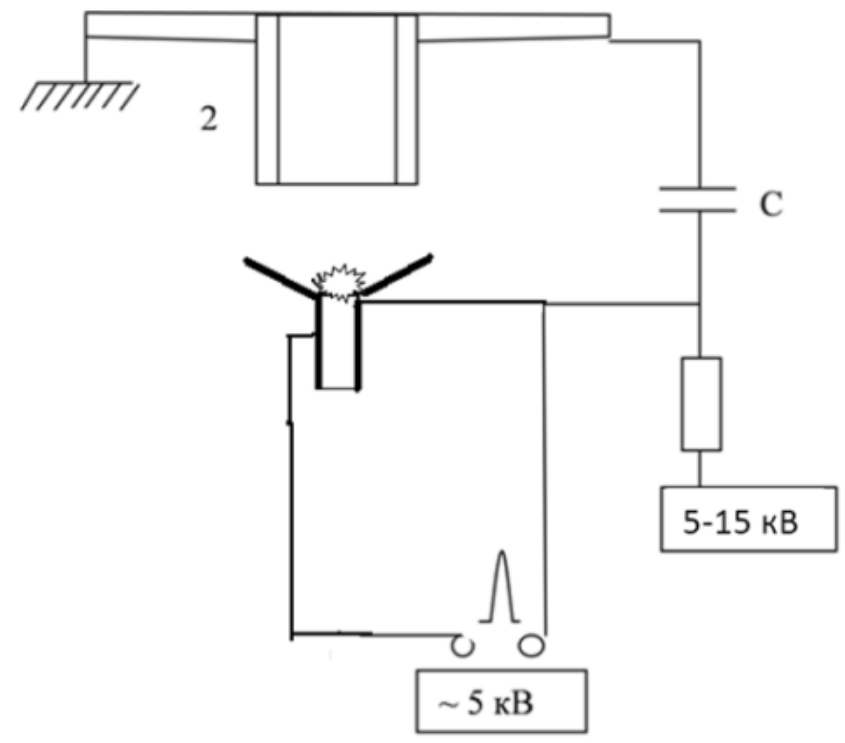

**Figure S17.** The design of the electron gun. 1 - cathode, 2 - anode.

A capacitor C was placed between the anode and cathode, charged from a high voltage source up to 5–15 kV. At the moment of breakdown, the energy of the capacitor *C* was expended on accelerating electrons from the breakdown plasma. Between pulses, capacitance C was recharged from a high-voltage source with a the time constant *RC*~1 ms. Note that the total duration of the current depended on the time of development and existence of the discharge plasma over the ferroelectric surface and the time of the discharge of the capacitor C through the plasma bunch formed as a result of breakdown over the surface. This time, in particular, depends on both the plasma parameters and the capacitance C and parasitic parameters determined by the design of the discharge gap.

### C. Control of parameters of e-beam

The pulsed e-beam was not strictly monochromatic, although most of the electrons had a maximum energy determined by the 12.5 kV voltage on the charged capacitor C. As this capacitor was discharged during the pulse time, the voltage across it and, accordingly, the intensity of the accelerating field decreased. In our experiments, a diaphragm with a typical diameter of 4 mm was installed in front of the sample, which limited the size of the irradiated region of the sample. To measure the current of the electron beam incident on the sample, we used a custom design developed for this purpose (see Fig.S18). The e-beam was formed in the plane of the sample using magnetic lenses and a deflecting system. It is known that the focal length of magnetic lenses depends on the energy of electrons. If the beam is not strictly monoenergetic, then electrons with different energies are focused on different distances. In this case, since the emissivity of the cathode increases with increasing accelerating voltage, most of the electrons have a maximum energy. When adjusting magnetic lenses by adjusting the current through them to obtain the maximum e-current, the latter is focused on the plane of the diaphragm in front of the sample. Electrons with lower energy are focused on other planes, and only a part of them flies through the diaphragm. Thus, the diaphragm in front of the sample is an energy selector of electrons: the smaller its size, the more monoenergetic the electron beam will be. Accordingly, the total duration of the e-beam current pulse in the case of a non-monoenergetic beam, will depend on the diameter of the diaphragm.

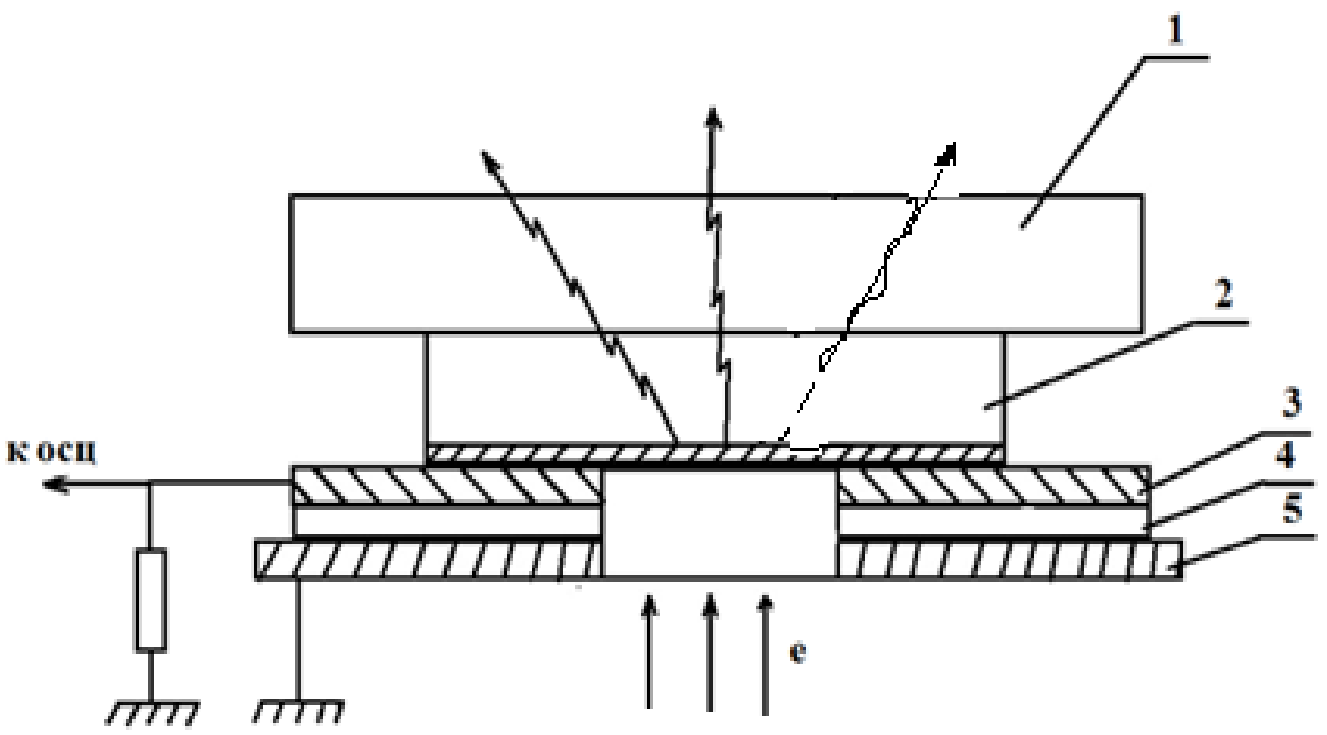

**Figure S18.** Scheme for measuring the electron beam current (pump current). 1 – sapphire window of the vacuum chamber, 2 – sample with a deposited Al-layer, 3 – diaphragm made of soft metal (indium), 4 – insulator (mica), 5 – copper diaphragm. Figure S19 shows three typical pulses of the e-beam current with a total pulse duration of 0.4-0.5 μs each, which consisted of an irregular sequence of series of short pulses with a duration from ~1 ns to tens of nanoseconds with a different shape.

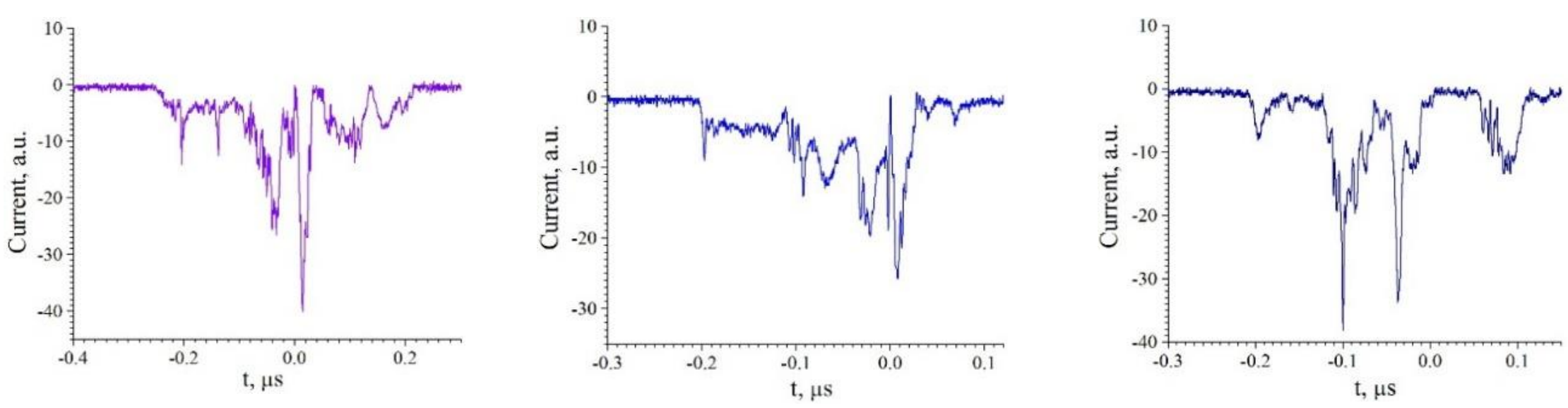

**Figure S19.** Typical e-beam current pulses measured for several discharges with a 4 mm diaphragm.

Apparently, the shape and magnitude of the current pulses depended both on the specific state of the discharge gap near the ferroelectric surface and on the magnitude of the magnetic field in the focusing and deflecting coils. Since the samples of GaN/AlN MQW structures were grown on *c*-sapphire dielectric substrates, this led to significant electric charging excited by an *e*-beam exacerbated by the high pulsed beam currents used (up to several amperes). To eliminate charging-induced measurement artifacts and prevent a potential dielectric breakdown of the samples, a thin conductive aluminum (Al) film was deposited on their surface. Due to the relatively high reflection coefficient of Al in the UVC range (>80%), this coating also reflected part of the UVC radiation from the structure towards the substrate.

The e-beam current passed through the diaphragm and absorbed by the sample passed through a 50 Ω resistor placed outside the vacuum chamber, the voltage pulse across which was measured by a high-frequency oscilloscope like to Tektronix TDS3000B or MSO 4104 with a bandwidth of 350 MHz or 1 GHz, respectively.

### D. Measurements of power characteristics of output UVC radiation

The sample under study was fixed directly on the exit window of the vacuum chamber with dimensions that ensured the minimum distance between the sample and the photodetector (~20 mm), as shown in Figure S16. In the work, we used a standard industrial photocell for recording short light pulses - a coaxial photocell FEK-22SPU with large-diameter photocathode, 40 mm (active area 12.6 $cm^2$).[S7] This element measured the optical power of light pulses in a wide solid angle and in the wide spectral range 220–650 nm with a time resolution of 0.5 ns. This photocell provides a sensitivity of 30 μA/lm in this spectral range with an upper limit of linearity of 18 A. For UVC-radiation with a wavelength of more than 270 nm, the sensitivity of the element is 17 mA/W, and in the range of 230–240 nm it slightly decreases to 15.5–16 mA/W respectively. Figure S19 shows the schematic design of this photodetector and optical transmission spectrum of the output sapphire window of the camera, which shows that its transmittance in the range of 230–240 nm is 70–75%.

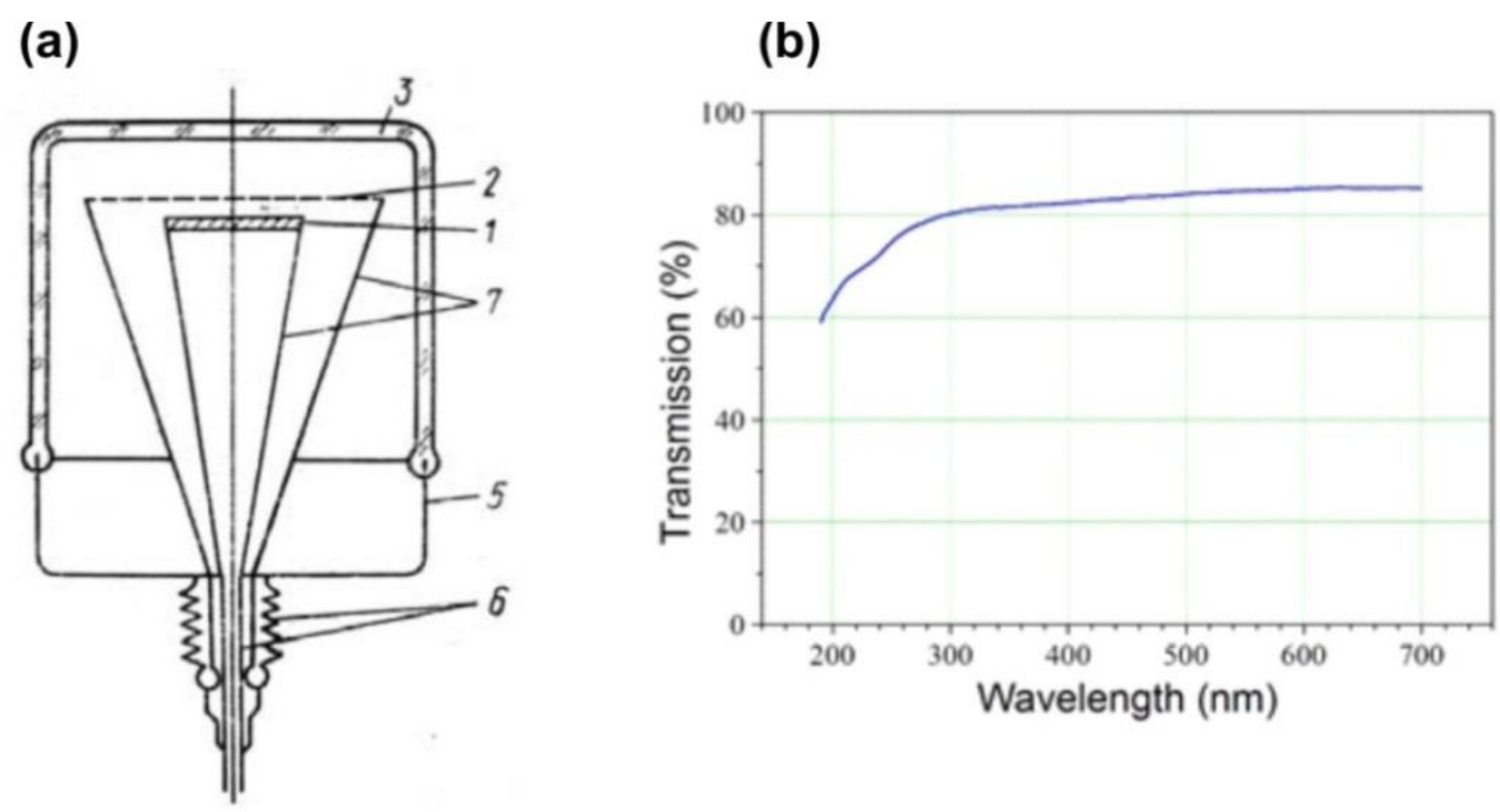

**Figure S20.** (a) Schematic design of photocell FEK-22SPU: *1* – photocathode, *2* - anode, *3* – input window, *5* – metallic body, *6*- coaxial connector, *7*- matching transition with equal impedance. (b) Optical transmission spectrum of the vacuum chamber window.

In addition, it should be taken into account that in the FEK-22SPU photocell, radiation enters the photocathode through the anode, made in the form of a "blind" - a set of thin metal plates, the plane of which is perpendicular to the flat surface of the photocathode. When measuring the spontaneous emission of structures, a significant fraction of the light passes at large angles to the cathode surface and part of the radiation falls on the side surfaces of the anode plates, being partially absorbed by them, and the radiation loss depends on the angle of incidence of light on the photodetector. This issue was considered in detail in our first paper in 2015 on electron-pumped UVC-emitters.[S8]

Thus, the absolute values of the output optical power of GaN/AlN MQW structures were calibrated as mA/W taking into account all above factors. In all our works since 2015 we use this type of photocell with a constant calibration.

**E. Meaurements of CL spectra excited by a high-current e-gun**

Figure 5a show the CL spectra of the *F*-series structures measured. Due to the non-reproducible nature of each *e*-beam pumping pulse from *e*-gun with a plasma ferroelectric cathode, it was impossible to record emission spectra using standard spectrographs based on conventional photomultipliers and stepper motors, which were used for the measurements of all CL spectra excited by low-current *e*-beam. To measure the CL emission spectra excited by a high-current *e*-beam we used a miniature S100 spectrograph with a concave diffraction grating,[S9] and a CCD array for detection. The device provided measurements in the spectral range of 190 – 1100 nm with a resolution of 1 – 2 nm. The signal accumulation time was usually 10 – 20 seconds at a pulse repetition rate of 12 Hz. At a lower frequency (1.5 Hz), recording was possible only in the presence of intense luminescence, since noise increased correspondingly with an increase in the accumulation time.